\documentclass[useAMS,usenatbib]{mn2e}

\usepackage{graphicx}
\usepackage{epsfig}

\newcommand{\Sauron}{\texttt{SAURON}}
\newcommand{\XSauron}{\texttt XSAURON}

\newcommand{\kms} {$\mbox{km s}^{-1}$}
\newcommand{\kmskpc} {$\mbox{km s}^{-1}\;\mbox{kpc}^{-1}$}

\newcommand{\Ha}{H$\alpha$}
\newcommand{\Hb}{H$\beta$}

\newcommand{\lda}{$\lambda$}

\newcommand{\OIII}{[{\sc O$\,$iii}]}
\newcommand{\NI}{[{\sc N$\,$i}]}

\def\etal{et al.~}
\def\Vrot{$V_{\rm rot}$}

\def\deg{^{\circ}}

\def\spose#1{\hbox to 0pt{#1\hss}}
\def\lta{\mathrel{\spose{\lower 3pt\hbox{$\sim$}}
    \raise 2.0pt\hbox{$<$}}}
\def\gta{\mathrel{\spose{\lower 3pt\hbox{$\sim$}}
    \raise 2.0pt\hbox{$>$}}}
\newdimen\hssize
\newdimen\hdsize
\newcommand{\Vsys}{$V_{\rm sys}$}
\newcommand{\Vlos}{$V_{\rm los}$}

\newcommand{\ten}[1] {10$^{#1}$}
\newcommand{\cmc}{cm$^{-3}$}
\newcommand{\ergscm}{erg s$^{-1}$ cm$^{-2}$}

\newcommand{\farcsec}{\hbox{$.\!\!^{\prime\prime}$}}

\newcommand{\NIwa}{\hbox{[N\,{\sc i}]$\lambda $5198}}
\newcommand{\NIwb}{\hbox{[N\,{\sc i}]$\lambda $5200}}
\newcommand{\NIww}{\hbox{[N\,{\sc i}]$\lambda\lambda $5198,5200}}

\newcommand{\OIIIwa}{\hbox{[O\,{\sc iii}]$\lambda $4959}}
\newcommand{\OIIIwb}{\hbox{[O\,{\sc iii}]$\lambda $5007}}
\newcommand{\OIIIww}{\hbox{[O\,{\sc iii}]$\lambda\lambda $4959,5007}}

\title[Gas and stellar dynamics in NGC~1068]{Gas and stellar dynamics in NGC~1068.\\
         Probing the galactic gravitational potential}
\author[Eric Emsellem \etal]{Eric Emsellem$^{1}$\thanks{E-mail:
emsellem@obs.univ-lyon1.fr}, Kambiz Fathi$^{2,3}$, Herv\'e Wozniak$^{1}$, Pierre Ferruit$^{1}$, 
\newauthor Carole G. Mundell$^{4}$, Eva Schinnerer$^{5}$ \\
$^{1}$ CRAL-Observatoire, 9 avenue Charles Andr\'e, 69231 Saint Genis Laval, France \\
$^{2}$ RIT Dept. of Physics, 84 Lomb Memorial Dr., Rochester, NY 14623-5603, USA \\
$^{3}$ Kapteyn Astronomical Institute, P.O. Box 800, 9700 AV Groningen, The Netherlands \\
$^{4}$ Astrophysics Research Institute, Liverpool John Moores University, 12 Quays House, Egerton Wharf, Birkenhead CH41 1LD, UK  \\
$^{5}$ Max-Planck-Institut f\"ur Astronomie, K\"onigstuhl 17, 69117 Heidelberg, Germany \\
}
\begin{document}

\date{Accepted 2005. Received 2005 ; in original form 2005}

\pagerange{\pageref{firstpage}--\pageref{lastpage}} \pubyear{2005}

\maketitle

\label{firstpage}

\begin{abstract}
We present \Sauron\ integral field spectrography of the central 1.5~kpc
of the nearby Seyfert~2 galaxy NGC~1068, encompassing the well-known  
near-infrared inner bar observed in the K band. We have successively disentangled the respective
contributions of the ionized gas and stars, thus deriving their two-dimensional
distribution and kinematics. The \OIII\ and \Hb\ emission lines exhibit very different spatial 
distribution and kinematics, the latter following inner spiral arms with clumps 
associated with star formation. Strong inwards streaming motions are observed 
in both the \Hb\ and \OIII\ kinematics. The stellar kinematics also exhibit clear signatures of a
non-axisymmetric tumbling potential, with a twist in both the velocity
and Gauss-Hermite $h_3$ fields. We re-examined the long-slit data of 
Shapiro \etal (2003) using a penalized pixel fitting routine: a strong
decoupling of the Gauss-Hermite term $h_3$ is revealed, and the central
decrease of Gauss-Hermite term $h_4$ hinted in the \Sauron\ data is confirmed. 
These data also suggest that NGC~1068 is a good candidate for a so-called 
$\sigma$-drop. We confirm the possible presence of two separate pattern speeds 
applying the Tremaine-Weinberg method to the Fabry-Perot \Ha\ map (Bland-Hawthorn \etal, 1991).
We also examine the stellar kinematics of bars formed in 
N-body $+$ SPH simulations built from axisymmetric initial
conditions approximating the luminosity distribution of NGC~1068. The resulting velocity, dispersion,
and higher order Gauss-Hermite moments successfully reproduce 
a number of properties observed in the two-dimensional kinematics of NGC~1068, and
the long-slit data, showing that the kinematic signature of the
NIR bar is imprinted in the stellar kinematics. The remaining differences 
between the models and the observed properties are likely mostly due to the exclusion of 
star formation and the lack of the primary large-scale oval/bar in the simulations.
These models nevertheless suggest that the inner bar could drive a significant amount of gas 
down to a scale of $\sim 300$~pc. This would be consistent with the
interpretation of the $\sigma$-drop in NGC~1068 being the result of central gas 
accretion followed by an episode of star formation.
\end{abstract}


\section{Introduction}

  The fueling of Active Galactic Nuclei (AGN) poses the problem of bringing gas
  in the close neighbourhood of the putative central dark mass, a supermassive
  black hole. Before reaching scales of a few parsecs where turbulent viscosity
  becomes important (Wada \& Norman 2002), the angular momentum of the gas must
  decrease by orders of magnitude.  QSO are usually associated with a major
  merging event which provides the necessary time varying potential to allow
  this to happen efficiently (Canalizo \& Stockton 2001). However, the presence of a Seyfert nucleus does
  not correlate significantly with the presence of companions
  or other external environmental properties (Maia et al. 2003). In this context, departures from
  axisymmetry in the gravitational potential have been advocated to play an important
  contribution in the removal of angular momentum of the dissipative component
  (Heller \& Shlosman 1994).

  Quadrupole perturbations such as bars are ubiquitous in disk galaxies.  And
  indeed bars are good at redistributing the gas, and more specifically at
  concentrating gas within their inner regions (Sakamoto et al. 1999).  However
  a correlation between the presence of a bar and the activity of the nuclear
  region is weak (Laine \etal 2002, 
  Knapen \etal 2000, Moles \etal 1995, Mulchaey \&
  Regan 1997, Malkan \etal 1998). This is not too surprising since the
  scales involved are very different: from the kiloparsec scale bars to the
  presumed central accretion disk. The bar-driven
  loss of angular momentum mainly occurs within the Corotation Radius (CR),
  leading the gas towards the inner resonances, e.g. the Inner Lindblad
  Resonance (ILR) if present, where the torques are cancelled out. Gas 
  accumulates at this radius, forming an inner ring in which clumps of
  vigorous star formation are often observed (e.g. Schwarz 1981, 1984).
  The next step in moving gas to smaller radii is still under great debate. A number of
  processes, including secondary inner bars, inner spirals, lopsidedness and
  minor mergers (see Combes 2003 for a review) have been invoked, but none appear
  to provide a necessary and sufficient condition for the triggering
  of nuclear activity.
	 
  Although gas is a very sensitive tracer of the presence of a barred
  potential (e.g. Mundell \& Shone 1999), its non-linear reponse to even
  weak non-axisymmetries means that it cannot be used directly to derive the
  gravitational potential of the galaxy. In addition, gas flows close to the
  AGN may be dominated by non-gravitational forces due to jets and outflow
  winds (Nelson \& Whittle 1996). 
  Therefore, the stellar kinematics, although challenging to
  measure in active galaxies, offer a more direct probe of the underlying
  potential.

  The difficulty in making reliable measurements of the stellar kinematics
  in AGN, particularly in nuclear regions complicated by the presence of
  strong line emission, has resulted in only a limited number of
  moderate-resolution, stellar absorption line studies. Two-dimensional
  stellar kinematics have been published for only a small number of Seyferts
  (e.g. Garc\'{\i}a-Lorenzo et al. 1997; Arribas et al. 1997;
  Arribas et al. 1999; Ferruit et al. 2004)
  and in some cases suffer from a too restricted field of view
  for accurate determination of the galaxy potential. For most Seyferts,
  only long slit studies (P\'erez et al. 2000;
  Emsellem et al. 2001; Filippenko \& Ho 2003; M{\' a}rquez et al. 2003) or central velocity dispersion
  measurements (Nelson \& Whittle 1995, 1996) are available.

  As one of the closest and most famous Seyfert~2 galaxies, NGC~1068 has been studied at most
  wavelengths and nuclear gas kinematics have been used to constrain its
  overall mass distribution. A good summarizing sketch
  of the observed components can be found in Schinnerer \etal (2000, their Fig.~6) 
  where the outer disk and oval,
  the two-arm spiral and the NIR bar are represented.  The outer oval structure
  (with a diameter of $90\arcsec$) has been interpreted 
  by Schinnerer \etal (2000) as a {\em primary} bar with
  $\Omega^P_p \sim 35$~\kmskpc, consistent with the HI ring being at its
  Outer Lindblad Resonance (OLR), and the inner spiral arms seen in CO
  corresponding to its Inner Lindblad Resonance (ILR).  The NIR bar, which
  extends up to a radius of $\sim 16\arcsec$ would then
  be a {\em secondary} bar with an estimated pattern speed of $\Omega^S_p \sim
  140$~\kmskpc. This is consistent with the recent lower limit estimate of $135 \pm
  42$~\kmskpc\ derived by Rand \& Wallin (2004) based on an application 
  of the Tremaine-Weinberg (1984) method using CO observations
  of the molecular gas. 
  The same authors argued for a different slower pattern speed for the 
  CO spiral arms with $\Omega^S_p \sim 72$~\kmskpc. However, based
  on the openess of the spiral arms, Yuan \& Kuo (1998) argued these to be
  associated with the OLR of the inner bar. The only previous two-dimensional stellar
  kinematics study was achieved by Garc\'{\i}a-Lorenzo  \etal (1997) who obtained
  {\tt INTEGRAL/WHT} spectroscopy in the optical of the central
  24\arcsec$\times$20\arcsec\ of NGC~1068. These authors suggest the presence of
  two puzzling kinematically decoupled stellar components in the central 10\arcsec,
  offset by about 2\farcs5 from each other. Finally, Shapiro \etal (2003) recently
  constrained the velocity dispersion ellipsoid of the disk using high-quality long-slit
  kinematics along the major and minor axes of the galaxy.
	 
  In this paper we wish to further constrain the gravitational potential of
  NGC~1068 by mapping both its stellar and gas kinematics. We have thus used the
  integral field spectrograph \Sauron\ to probe the region of the inner
  (secondary) bar. We report the results obtained from this
  dataset, analysed with the help of N-body $+$ SPH simulations.  We first
  describe the \Sauron\ observations and associated data reduction in
  Sect.~\ref{sec:data}. We then briefly present additional data we have gathered
  from different authors (Sect.~\ref{sec:otherdata}), and present the results
  in Sect.~\ref{sec:results}. Numerical simulations for NGC~1068 are described in
  Sect.~\ref{sec:dynmodel}, and the resulting kinematics are then compared with our
  data. The results are discussed further in Sect.~\ref{sec:discuss}, and
  summarized in Sect.~\ref{sec:conclude}. Throughout this paper, 
  we will use a distance of 14.4~Mpc for NGC~1068 (Bland-Hawthorn \etal 1997), 
  leading to a scale of 69.8~pc per arcsecond.
  
\section{\Sauron\ observations and data reduction}
\label{sec:data}

\subsection{The \Sauron\ datacubes}

We have observed NGC~1068 with the Low Resolution (LR) mode
of the integral field spectrograph \Sauron\ during a run
in January 2002. \Sauron\ delivers about 1500 spectra simultaneously, 
homogeneously covering a field of view of $41\arcsec\times33\arcsec$ 
with a squared sampling of 0\farcs94 per spatial element (lens).
The spectral domain and resolution are [4820 - 5280 \AA] and 108~\kms\ ($\sigma_{\star}$) respectively.
More details on the \Sauron\ spectrograph can be found in Bacon \etal\ (2001),
de Zeeuw~\etal\ (2002), and Emsellem~\etal\ (2004).
The emission lines in NGC~1068 (particularly the \OIII\ lines) are so bright in the central
part of NGC~1068 that the CCD saturates after about 8 min of exposure with the \Sauron\ LR mode.
We have therefore obtained a 5 min exposure to probe the central few arcseconds,
and three more exposures of 30 min to reach sufficient a signal-to-noise ratio at the edge of the field of view.

\subsection{Data reduction with \XSauron}

The data reduction of all four exposures was achieved using the dedicated \XSauron\ software, and
an automated pipeline available within the \Sauron\ consortium (see Bacon \etal 2001, and de Zeeuw
\etal 2002 for details). The main steps include: removal of the CCD signature, extraction of the
spectra using a mask built from an optical model of the telescope$+$spectrograph, wavelength
calibration, low frequency spectral fielding, cosmic rays removal, homogenization of the spectral
resolution in the field, subtraction of the sky contribution using 146 dedicated lenses
(1.9\arcmin\ away from the central field), and flux calibration. The sky spectra
were carefully checked for any significant contribution (emission/absorption) from NGC~1068 itself.

The flux-calibrated individual exposures are then accurately centred relatively
to each other using the associated reconstructed images, 
and merged. Since the exposures are slightly offset from each other, we use
the dithered datacubes to spatially resample the datacubes to 0\farcs8 per spaxel (spatial pixel).
Before merging, the spectra which exhibited saturated pixels were removed from the datacubes.
This practically means that the spectra in the central 4\arcsec\ of the final merged datacube are
only coming from the 5 minutes unsaturated exposure. At the edge of the \Sauron\
field of view, the signal-to-noise ratio gets down to about 30 per pixel, sufficient
for our goal of deriving the stellar kinematics.

\subsection{Coordinate centering}
\begin{figure}
\centering
\epsfig{file=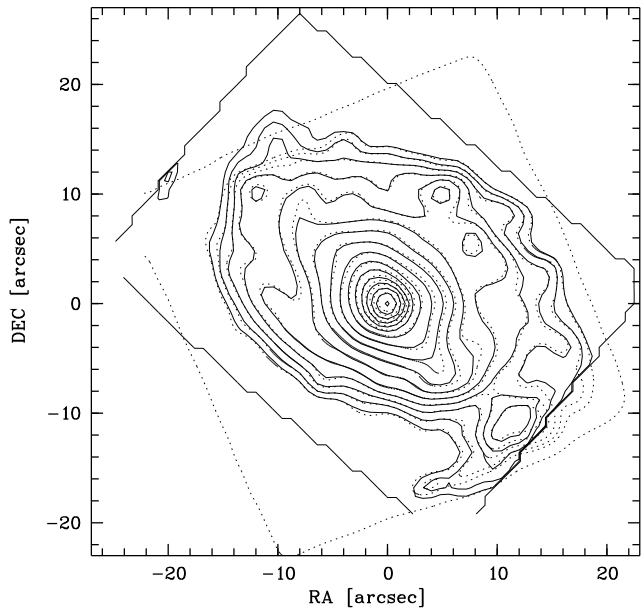, width=\columnwidth}
\caption{Isophotes (with steps of 0.5 magnitude) of the central region as
	observed by \Sauron\ (stellar continuum, solid lines) and with the WFPC2 camera aboard HST
		(F547M filter, dashed lines),
	after (seeing-)convolution by a gaussian of 2\arcsec. The agreement is
		excellent except at the border of the HST/PC field where edge effects due
      to the convolution are seen.}
\label{fig:hst}
\end{figure}

We used the F547M WPFC2/HST (extracted from the ESO/ST-ECF archive) 
data to determine the relative spatial position of our
\Sauron\ datacube (see Fig.~\ref{fig:hst}). Capetti \etal (1997) provided the absolute coordinate of the
central peak of this optical image ($\alpha$ = 02h42m40.s711, $\delta$ = -00deg00'47."81 
- J2000, FK5, 80 ~mas accuracy), to be compared with the putative position of the central engine
(CE) identified as the S1 radio source by Muxlow \etal (1997) at $\alpha$ = 02h42m40.s7098, $\delta$ = -00deg00'47."938. 
Taking S1 as our reference (0,0), we thus aligned the \Sauron\ reconstructed
image with the F547M WFPC2 exposure where the peak was assumed to be 0\farcs13
North and 0\farcs02 East of the CE (including a potential rotation of the
\Sauron\ field of view). This ensures an absolute positioning of
our \Sauron\ datacube on S1 with an error better than 0\farcs1 in translation and
than $0.5\deg$ in rotation (the latter being dominated by the uncertainty on the
relative angle between the  WFPC2 and \Sauron\ data). 
All maps presented in this paper are oriented in the classical way, with North up, and East left.

\subsection{Stellar kinematics}

The stellar contribution to the \Sauron\ spectra is highly contaminated by 
strong emission lines: \Hb\lda4861, the \OIII\lda4959,\lda5007 and \NI\lda5198,\lda5200
doublets. We first identified the spectral regions which are significantly contaminated
by emission. This required a detailed examination of the spectra in
the merged datacube, particularly in the central 5\arcsec\ where the emission
lines are very strong and wide. A first estimate of the stellar kinematics is then 
derived excluding the contaminated pixels. We achieve this by using a direct
penalized pixel fitting routine (pPXF, Cappellari \& Emsellem 2004): 
the algorithm finds the mean velocity $V$ and velocity dispersion $\sigma$ which
minimizes the difference between the observed galaxy spectrum and the spectrum
of a stellar template convolved by the corresponding gaussian (of mean $V$ and
root mean square $\sigma$). This is performed with the galaxy and template star 
spectra rebinned in $\ln\lambda$.
We use a single template star for this first estimate, namely the K2 giant HD~26162,
also observed with \Sauron.
Best-fit values for $V$ and $\sigma$ are thus obtained at each position (for each
lens/spectrum) independently. 

\begin{figure*}
\centering
\epsfig{file=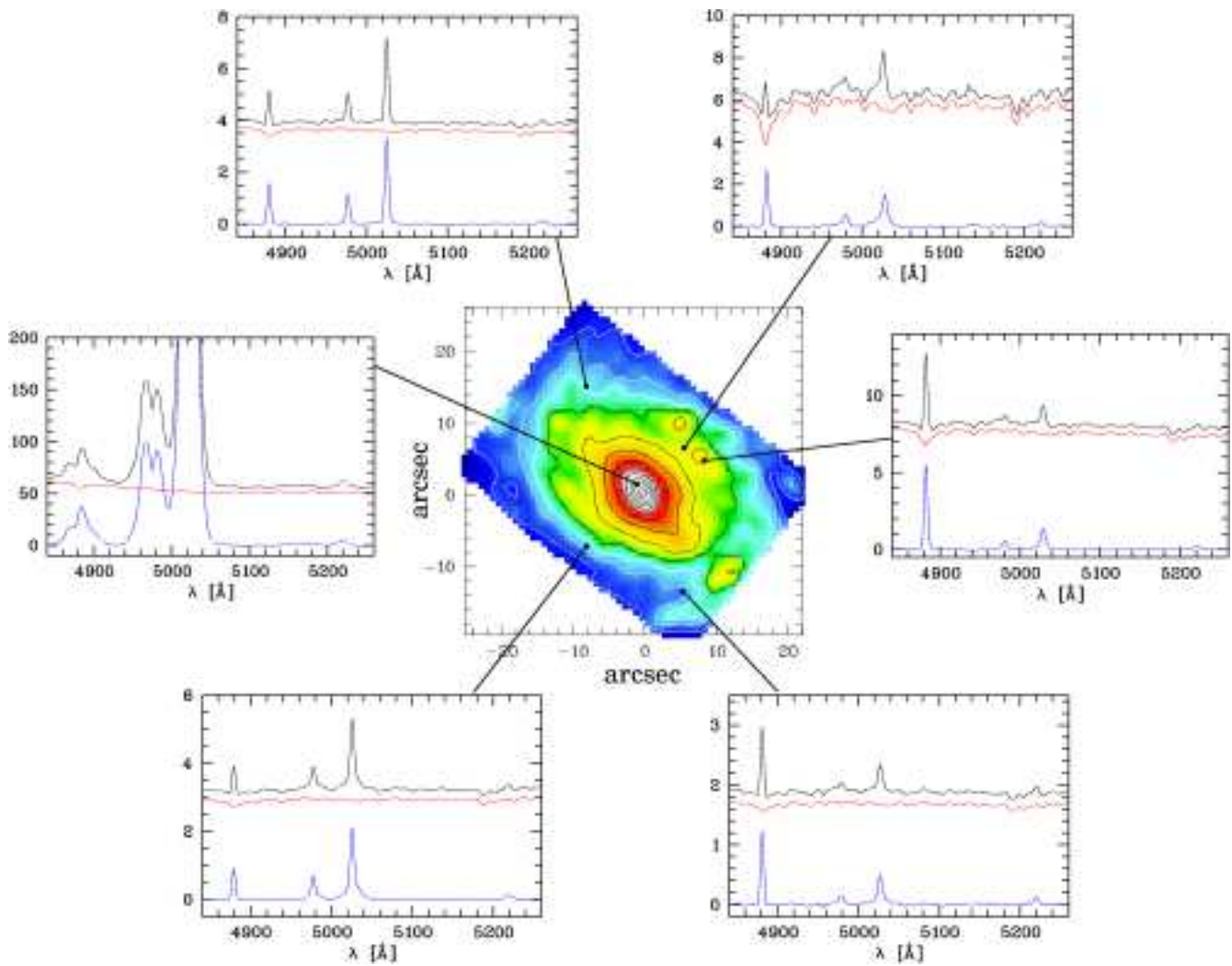, width=\hdsize}
\caption{Fits of the \Sauron\ spectra of NGC~1068 using the Vazdekis stellar library.
Six spectra (black lines) and their corresponding fits (red lines) 
at different locations in the \Sauron\ field are shown,
the individual spatial locations being indicated by arrows on the reconstructed intensity \Sauron\
map. The vertical position of the best-fit spectra are arbitrary shifted for legibility, the residual
spectra being presented in the same panels (blue lines). The \Sauron\ map has North up, and
East left.}
\label{fig:ssp}
\end{figure*}

We then use this initial estimate of the stellar kinematics to derive an optimal
stellar template for each individual spectrum. This is usually done using
a large library of stellar templates. In the case of NGC~1068, and because 
of the significant number of pixels we had to exclude from the fit, this would
make the fitting process strongly degenerate. 
We therefore decided to restrict our stellar library to
include three different stellar templates 
from the single-age single-metallicity stellar population (SSP) models of Vazdekis (1999);
this proved sufficient to properly fit the underlying stellar contribution (see Fig.~\ref{fig:ssp}).
A multiplicative polynomial with a maximum degree of 6 was included in the fit
in order to account for small residual differences between the stellar libraries
and the \Sauron\ spectra.

We finally iterate by measuring the velocity $V$ and dispersion $\sigma$, as
well as the third and fourth Gauss-Hermite moments $h_3$ and $h_4$
with the pPXF routine, this time using the optimal templates obtained from the
previous step. $h_3$ and $h_4$ are indicators of the skewness and peakiness of
the line-of-sight velocity distribution (relatively to a Gaussian). 
Although the signal-to-noise ratio per pixel ($> 30$) is sufficient everywhere in the
\Sauron\ field of view to derive reliable $h_3$ and $h_4$ values, these should
be taken with caution considering the limited spectral domain of the \Sauron\ datacube, the medium
spectral resolution of the \Sauron\ spectra ($\sigma = 108$~\kms) and more
importantly the contamination by bright emission lines. A comparison with
long-slit data is helpful in this context (see Sect.~\ref{sec:skin}).

\subsection{Gas distribution and kinematics}
\label{SectionELfitting}
\begin{figure*}
\epsfig{file=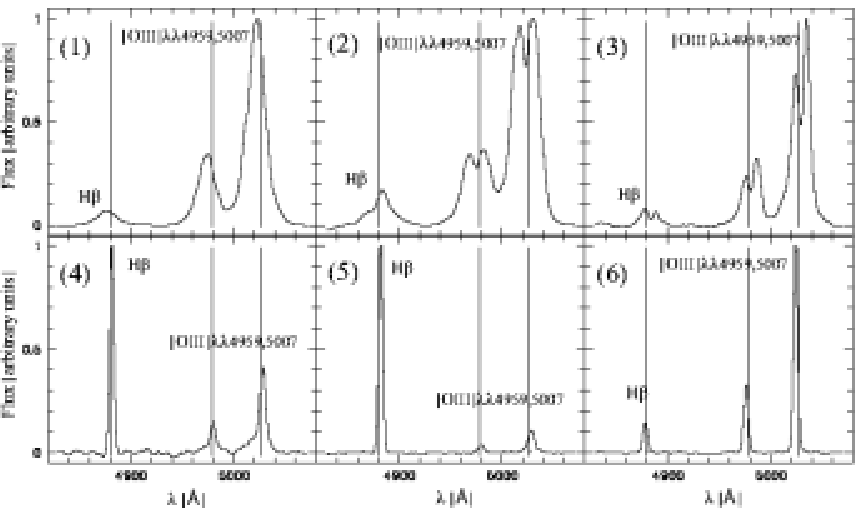, width=\hdsize}
\caption{Examples of spectra of NGC~1068 obtained with \Sauron. The spectra,
	normalized to a peak intensity of unity, have been truncated to show only
		the wavelength range including the \Hb\ and \OIIIww\ lines. The dotted
		vertical lines indicate the expected position of a line at rest in the
		galaxy frame (systemic velocity of 1144~\kms). (1) Nuclear spectrum. (2)
      and (3) Spectra located at positions ($-$1\farcsec3,$+$1\farcsec2) and 
      ($+$0\farcsec5,$-$2\farcsec8)
      where both the narrow and additional components are present. In panel 2
      the narrow component is redshifted with respect to the systemic velocity
      (the additional component being blueshifted), 
      whereas in panel 3 it is the opposite.
      (4) Spectrum located at
		($+$5\farcsec3,$+$8\farcsec4) in which both the narrow and broad
		components are present in the \OIII\ lines. (5) and (6) Spectra located
		($+$13\farcsec3,$-$6\farcsec8) and ($-$16\farcsec3,$+$14\farcsec8),
	respectively and in which only the narrow component is observed (but with
			very different \OIIIwa\ / \Hb\ ratios).\label{fig:FigureSpectra} }
\end{figure*}

The spectra resulting from the fitting procedure described in the previous
Section were subtracted from the original data to provide pure
emission lines spectra. The wavelength range of our 
observations includes the \Hb, \OIIIww\ and \NIww\
emission lines. These five lines are detected over our complete field of
view, although the lines of the \NI\ doublet were significantly weaker than
the \Hb\ and \OIII\ lines and were barely detected in off-nuclear regions.
The basic parameters of these emission lines (intensity, centroid velocity
and velocity dispersion) were derived from Gaussian fitting of their profile using the
{\it fit/spec} software (Rousset 1992). When relevant, the lines were
constrained to have fixed or bounded ratios (\OIIIwb\ / \OIIIwa\ = 2.88; 0.7 $<$
\NIwa\ / \NIwb\ $<$ 2.0)\footnote{Range estimated using the Mappings Ic
software with an electronic temperature of \ten{4}~K
and electronic densities from 0.1~\cmc\ to 1000~\cmc\ (Ferruit et al. 1997).}. To stabilise the
fit of the weak \NI\ lines, we forced them to share the same
velocity and width as the \Hb\ line. 

Careful examination of the spectra showed the presence of several kinematically
distinct components in the profiles of the \Hb\ and \OIII\ emission lines, in
agreement with the results of Arribas et al. (1996) and
Garc\'\i a-Lorenzo et al. (1999). In our data, we have
identified 3 different systems: 
\begin{itemize} 
\item a first component, which is observed everywhere in our field of view. It is relatively narrow 
(typical dispersion of 100~\kms\ or lower beyond 5\arcsec\ from the nucleus)
everywhere except in the nuclear regions, and it is therefore termed
"narrow" hereafter. Despite its broadening in the nuclear region, and the presence of 
the other systems, it is possible to follow this "narrow" component down to
about 2\arcsec\ from the centre.
\item a broad (dispersion higher than 600~\kms), 
blueshifted component (hereafter termed "broad"), which
is observed up to 8\arcsec-10\arcsec\ from the nucleus and often appears as a
blue wing in the \OIII\ line profile (see e.g., panel 4 of Fig.~\ref{fig:FigureSpectra}).
\item a very spatially localized, intermediate-dispersion component (hereafter
termed "additional") observed in the vicinity of the nucleus and either
strongly blueshifted (North-East of the nucleus) or redshifted (South-West of
the nucleus, see e.g., panels 2 and 3 of Fig.~\ref{fig:FigureSpectra}) with respect to the systemic
velocity of the galaxy.
\end{itemize}

Examples of spectra displaying these three kinematic components are shown in
Fig.~\ref{fig:FigureSpectra}, which can be compared with Fig.~7 in 
Garc\'\i a-Lorenzo et al. (1999). 
Our narrow, broad and additional components correspond
to components (1), (2)+(3) and (4a)+(4b), respectively, in 
Garc\'\i a-Lorenzo et al. (1999). In our data,
it was not possible to disentangle their components 2 and 3. This is probably due
to our spatial sampling (0\farcsec8 per lens), which makes it
difficult to study the (complex) central few arcseconds where component 3 is
identified by these authors. For the same reasons, conducting a similar
comparison with the decomposition used by Arribas et al. (1996) 
proved very difficult because their observations cover only the nuclear
regions (see their Sect. 3.2.1). It must also be emphasized that each system
identified in the \Sauron\ spectra may itself result from a blend of separate (and unresolved)
velocity systems: this is for instance clearly the case in the central few
arcseconds (Cecil \etal 2002).

At distances $>$~5\arcsec\ from the nucleus (corresponding to 1822 spectra out
of a total of 1943), the line profiles were simple enough for an automated
fit to be conducted. The various maps inferred from the results of this
automated fit were carefully checked and only a small number of spectra had to
be fitted a second time individually. In contrast, the automated fit to spectra in the
inner regions (radii less than 5\arcsec) was unstable, as expected. All spectra had to
be examined and fitted individually and quite often additional constraints
(especially on the width of the broad and additional components) were imposed. 
Our results in these inner regions (which, however, represent only a small
fraction of our field of view) are therefore less reliable than those for
the outer regions. 

Given the known kinematic complexity of the nuclear regions (Cecil, Bland \&
Tully 1990) and the limitations of our
dataset in this region, we do not discuss the properties of the
additional and broad components further. In the following, we focus on
the properties of the narrow component, which is more representative of
the underlying galaxy.

Errors on the emission line parameters were derived 
by use of Monte Carlo realizations repeating the fitting procedure 500 times 
using simulated emission line spectra. These were built as the
sum of a synthetic noise-free spectrum (with spectral characteristics typical
of the observed spectra) and noise. Additional errors on the centroid
velocities (18~\kms\ for 3$\sigma$ calibration errors) and the Full Width at
Half Maximum (FWHM, 0.1~\AA\
peak to valley for variations of the "instrumental" FWHM over the field
of view) were included in the final error budget. In the case of
the centroid velocities of the narrow component,  the contribution of the fit to the total error was
negligible for spectra dominated by this component (i.e. typically at radii $>$~5\arcsec)
and with a peak signal to noise ratio larger than 10 (i.e. over most of our
field of view): the overall accuracy on the measured centroid velocity is then
better than 20~\kms.
\begin{figure*}
\epsfig{file=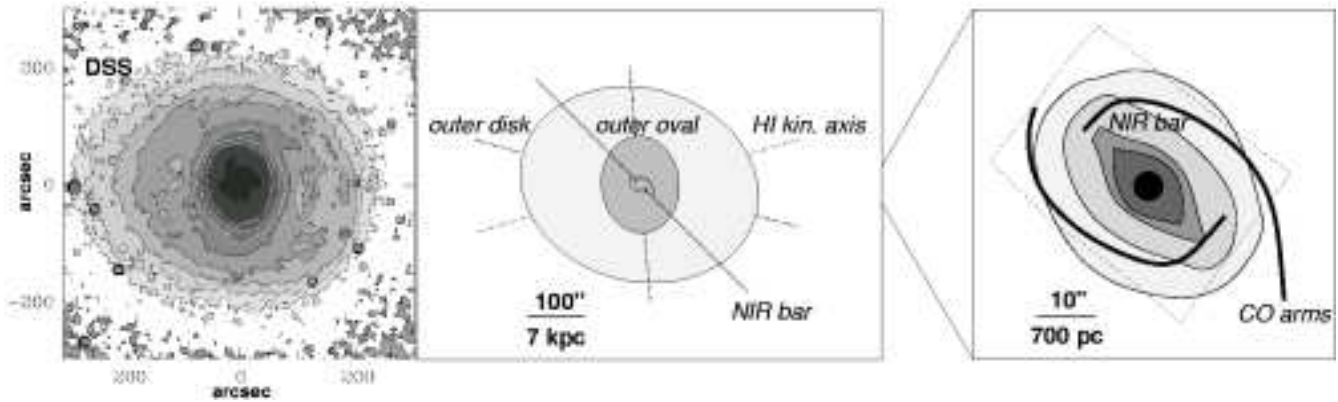, width=\hdsize}
\caption{Schematic of the structures observed in NGC~1068. 
Left panel: 10\arcmin$\times$10\arcmin\ DSS I band
image. The central 40\arcsec\ are saturated. Isophotes are shown with a
step of 0.5 mag.arcsec$^2$. Middle panel: sketch at the same scale as the DSS
image. The major-axes of the outer disk, outer oval and inner near-infrared bar
are indicated (dashed lines). The HI kinematic axis is shown as a dotted line
for comparison. Right panel: zoomed sketch ($\times 10$) showing the location of
the near-infrared bar and the CO arms (see Schinnerer et al. 2000, Fig.~1 and 6). The
extent of the \Sauron\ field of view is indicated by a dashed polygon.}
\label{fig:sketch}
\end{figure*}

\section{Ancillary datasets}
\label{sec:otherdata}

In this Section, we briefly describe ancillary datasets we use in this work, namely
some optical and near-infrared images as well as the HI velocity rotation curve for 
the dynamical modelling (Sect.~\ref{sec:dynmodel}), \Ha, and 
CO distribution and kinematics for comparison with our emission line \Sauron\ maps.

\subsection{Ground-based photometry}
\label{sec:photom}

A deep B band image was used to probe the outer disk of NGC~1068 up to a radius of 200\arcsec.
This image was obtained with the 1m telescope on Mount Laguna (Cheng \etal 1997)  and is a
combination of 3 exposures of 300s each. It has a field of view of about 800\arcsec\ sampled at 0\farcs4 per
pixel. We also made use of the Digitized Sky Survey I band (available via the ESO/ST-ECF 
archive: http://archive.eso.org/) and 2MASS K band images, 
both having a scale of about 1\arcsec\ per pixel. 
Finally, a high resolution K band image was obtained by Peletier \etal
(1999): this has a pixel size of 0\farcs248, a field of view of about 
1\arcmin$^2$, and a seeing of 0\farcs5 (FWHM).

\subsection{HI velocity curve}

The HI velocity curve we used has been published in the Ringberg Standards (RS hereafter,
Bland-Hawthorn \etal 1997) from the work of Brinks \etal (1997). It comprises measurements up
from the centre to about 200\arcsec, the last radius at which Brinks \etal (1997) detected the 
low surface brightness HI disk. The HI rotation curve decreases from about 130~\kms\ at 30\arcsec\
to 95~\kms\ at 180\arcsec\ from the centre.

\subsection{The \Ha\ distribution and kinematics}

Dr. J. Bland-Hawthorn provided us with the \Ha\ Fabry Perot data as published in Bland-Hawthorn \etal
(1991), including the luminosity and mean velocity maps. 
The datacube was obtained with the Hawaii Fabry-Perot Interferometer (HIFI) at the CFHT
(Mauna Kea). The velocity resolution was 65~\kms, the spatial sampling and resolution (FWHM) being
0\farcs43 and 0\farcs8 respectively. The \Ha\ velocity field was published
in Dehnen \etal (1997).

\subsection{The CO distribution and kinematics}

We also make use of the high resolution CO interferometric data published in Schinnerer
\etal (2000, S+00 hereafter). This $^{12}$CO(1-0) dataset has been obtained with
the IRAM millimeter interferometer on the Plateau de Bure (France), providing a circular beam of
1\farcs4 (sampling of 0\farcs4), with a channel width of 10~\kms.  Apart from a central ring-like structure,
the molecular gas is mainly distributed along a two-arm spiral just outside the inner
near-infrared bar, and which is associated with the ILR of the outer oval (S+00). 

\subsection{Long-slit stellar absorption kinematics}
\begin{figure*}
\centering
\epsfig{file=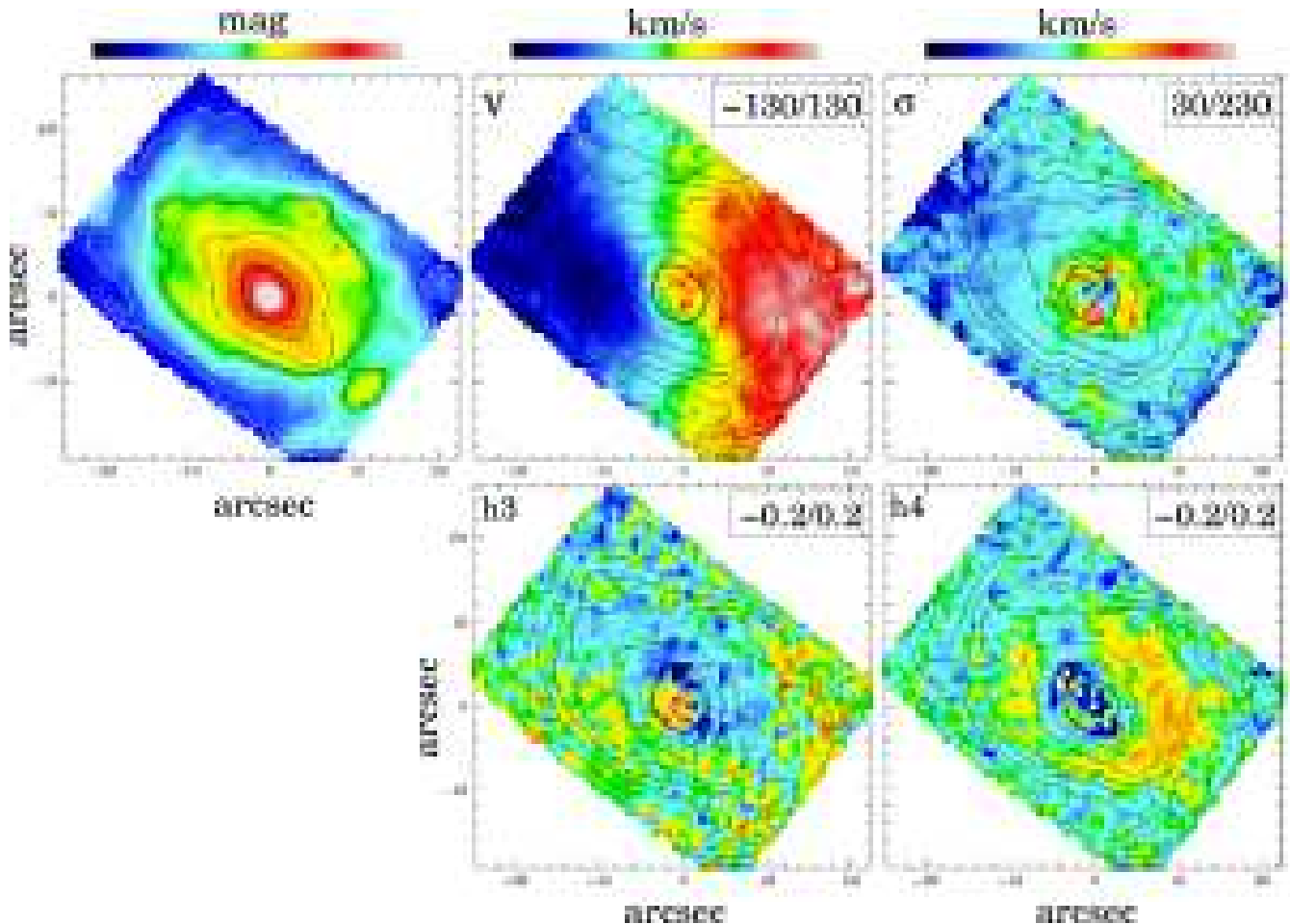, width=\hdsize}
\caption{Reconstructed maps of the stellar kinematics of NGC~1068 obtained from the \Sauron\ data cube.
From left to right: Intensity, velocity $V$, velocity dispersion
$\sigma$ (top panels), and higher order moments $h_3$, and $h_4$ (bottom panels). 
The thick contour approximately marks where the kinematic measurements are very
significantly perturbed by the presence of very strong and wide emission lines.
Boxes in the top right corner of the panels show the ranges covered by the color bars.}
\label{fig:starkinmaps}
\end{figure*}

Finally, we include results from long-slit spectroscopy conducted by Shapiro \etal (2003) who
kindly made the fully reduced datasets available to us: stellar kinematic
profiles along the major and minor-axes of the galaxy (PAs of 80 and 170 degrees)
were obtained with a slit width of 3\arcsec\ centred on the Mgb triplet around 5175\AA. 
The stellar kinematics published in Shapiro \etal (2003, hereafter Sh+03) were obtained using a Fourier fitting 
algorithm. In this paper, we reanalysed these data favouring the pPXF routine 
as it allows an optimal selection of regions uncontaminated by emission lines, 
and to derive robust estimates of moments up to $h_3$ and $h_4$. We have 
only included wavelengths between 5237 and 5549\AA\ to avoid 
contamination from the very bright emission lines present in these spectra of
NGC~1068. 
\section{Data analysis}
\label{sec:results}

\subsection{Morphology and position angles}

In this Section, we derive new values for the main 
morphological parameters of the different components
of NGC~1068 (bar, oval, outer disk), since these are important parameters when
comparing different datasets and models. A schematic summarizing the structures
observed in NGC~1068 is provided in Fig.~\ref{fig:sketch}.

We first wish to reexamine the position angle (PA) and ellipticity of the 
outer disk. The RS quote a value of
$106\deg \pm 5$, which in fact corresponds to the average position angle of
the {\em kinematic} axis as fitted on the large scale HI data (Brinks \etal 1997).
The total HI surface brightness in the outer disk (100--200\arcsec) is rather low
and exhibits a North/South asymmetry at its outer edge
(lower surface brightness in the South). At this radius the stellar
component is easily observed using the DSS I band image: outside 190\arcsec,
there is a clear mildly flattened component with a PA between 74.5 and $84\deg$,
therefore at least $20\deg$ from the kinematic PA measured in HI. A value
of $80\deg\pm5$ is consistent with the optical and smoothed HI surface
brightness shown in Dehnen \etal (1997), and we will adopt this value
as the apparent photometric PA of the outer disk, thus different from the HI kinematic PA. 
The axis ratio of the best fit ellipse of the optical outer disk is between 0.8 and 0.85.

We then remeasured the characteristics of the outer oval and near-infrared bar
using ellipse-fitting. Both the DSS I band and the B deep images lead to a radius of 90\arcsec,
an axis ratio of 0.8 and a PA of $5\deg$ for the outer oval, perfectly
consistent with the RS values. Using our high resolution K band image, 
we find an average position angle of $44.5\deg \pm 0.5\deg$ for the 
near-infrared bar (between 10 and 16\arcsec\ radius), as compared to the 
RS value of $48\deg \pm 2\deg$ (from Scoville \etal 1988). 
Our value is however consistent with the one
provided by Thronson \etal (1989) of $45.0\deg \pm 0.5\deg$. In the following,
we will therefore use an average value of $44.5\deg$. 
The minimum axis ratio is 0.45 at a radius of 15\farcs5. 

\subsection{Stellar kinematics}
\label{sec:skin}
The \Sauron\ stellar kinematics are presented in Fig.~\ref{fig:starkinmaps}.
Inside the central 4\arcsec, the measurements are significantly perturbed
due to the emission line contamination and should be taken only as indicative.
The stellar velocity field clearly exhibits strong departures from axisymmetry,
with an "S" shaped zero velocity curve, and the line of maximum velocity
having a changing position angle. The amplitude of the velocity field reaches 
$\sim 115$~\kms\ at a radius of about 10\arcsec. We wish to draw attention to 
a small perturbation of the order of 20\kms at the South-East edge of the
field (absolute values being smaller, around a band going from 12\arcsec\ East, 6\arcsec\ South
to -18\arcsec\ East from the nucleus; see Sect.~\ref{sec:gas}) and a similar
trend is mirrored on the opposite side. The velocity dispersion map shows a 
rise towards the center with $\sigma \sim 60$~\kms\ at 20\arcsec\ from the centre along the
major-axis and between 100 and 200~\kms\ in the central 10\arcsec. There is a
slight asymmetry in the dispersion map (dispersions being higher by $\sim 20$~\kms\
on the western side of the field), probably the result of a residual gradient of the spectral
resolution over the field of view: however, this does not affect our conclusions.
The $h_3$ map displays a significant change of sign from the SE quadrant to the NW quadrant. 
The structure of positive and negative $h_3$ is roughly elongated along the position
angle of the NIR bar. There is an elongated ring-like region of positive values in the
$h_4$ map (with maxima around 0.06--0.1 between radii of 8 and 12\arcsec) and a depression inside 
a radius $R$ of 8\arcsec\ with $h_4$ going slightly negative. We cannot however
follow this inward decrease of $h_4$ for $R < 4\arcsec$ because of the emission-line contamination.
\begin{figure}
\epsfig{file=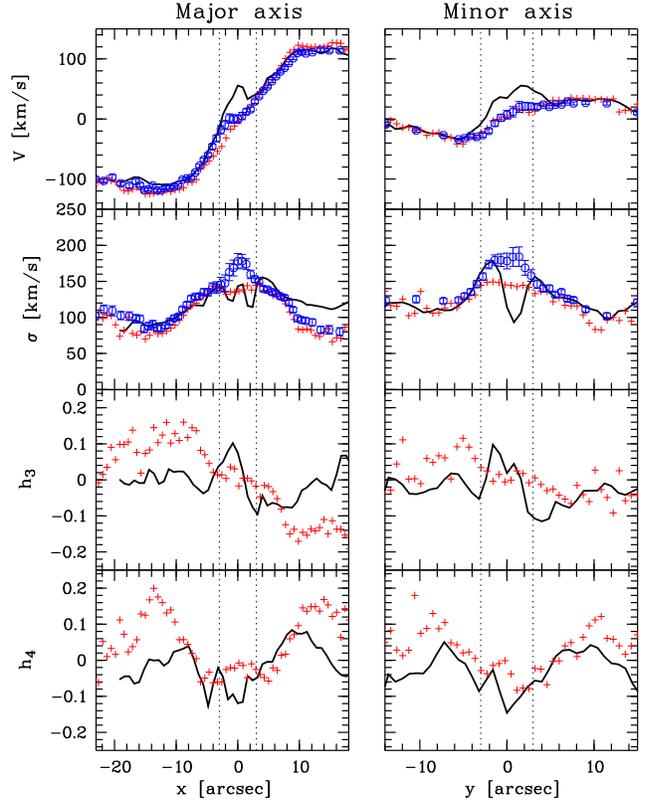, width=\columnwidth}
\caption{Comparison between stellar kinematic profiles extracted from the
\Sauron\ (solid lines) and Sh+03 (crosses). 
From top to bottom: mean velocity, velocity dispersion, $h_3$ and $h_4$. The \Sauron\ data
have been averaged over a slit width of 3\farcs2\ (the central 3\arcsec\ are unreliable
due to the presence of dominating emission lines). 
The mean velocity and dispersion
profiles originally published by Sh+03 are provided as circles
(with error bars). The "major-axis" here corresponds to a PA$=80\degr$, 
and the "minor-axis" to PA$=170\degr$.}
\label{fig:shapiro}
\end{figure}

We now compare the kinematics published by Sh+03 and obtained via 
a Fourier fitting technique, with our reanalysis of the same dataset using pPXF
(Fig.~\ref{fig:shapiro}). 
The central velocity dispersion does not peak so clearly in
the profiles reextracted from the Sh+03 data: we were unable to reproduce the
high central $\sigma$ value even by changing our stellar template or spectral domain.
There is also a slight flattening of the major-axis velocity gradient in the central 
2\arcsec\ in the Sh+03 profiles, which we do not see in our measurement. 
The $h_3$ profiles are anticorrelated with $V$ with peak values of $\pm 0.15$
around 10\arcsec\ along the major-axis. The $h_4$ profiles exhibit a significant
depression in the central 5\arcsec\ with a negative minimum of $\sim -0.04$ at the centre.
The maximum value of $h_4$ is reached at a radius of about 14\arcsec\ along the
major-axis and 10\arcsec\ along the minor-axis confirming the elongation of this
ring-like structure. The presence of a drop in the central dispersion profile 
is confirmed by Shapiro \& Gerssen (private communication) who reexamined
this dataset and applied their own pixel fitting routine. We presume that the central peak in the stellar
velocity dispersion observed in Sh+03 is due to the influence of 
the strong [NI] doublet when using a Fourier fitting program to extract the kinematics. 
Considering this reanalysis, NGC~1068 is
therefore a new candidate for the presence of a so-called $\sigma$-drop
(Emsellem \etal 2001; Wozniak \etal 2003; M{\' a}rquez \etal 2003).

Overall the \Sauron\ stellar kinematics averaged over a reconstructed slit of 3\farcs2\ width
compare reasonably well with the published long-slit data
of Sh+03, although there are significant discrepancies worth
mentioning: the \Sauron\ dispersion values are too high on the western side of
the major-axis, confirming the fact that the \Sauron\ data have a 
residual spatial variation of the spectral resolution. 
In contrast to the profiles measured from the Sh+03 data, our $h_3$ data do not
show a significant gradient. This is primarily due 
to our lower spectral resolution ($\sigma = 108$~\kms, as compared to about 30~\kms\
for Sh+03), which obviously also affects our $h_4$ measurements.

We do not see any hint for the decoupled kinematic
structure claimed by Garc\'{i}a-Lorenzo \etal (1999, 1997). 
Although the derived stellar kinematics within the very central
4\arcsec\ are unreliable, we should be able to detect the presence of 
a large East-West velocity gradient as implied 
by the pinched isovelocities in the maps of Garc\'{i}a-Lorenzo \etal (1997). 
There is also no hint of an abrupt velocity change in the minor-axis kinematics
of Sh+03. This strongly suggests that the kinematic structure
observed by Garc\'{i}a-Lorenzo \etal (1999, 1997) is an artefact in the \texttt{INTEGRAL}/WHT
data.
\begin{figure*}
\epsfig{file=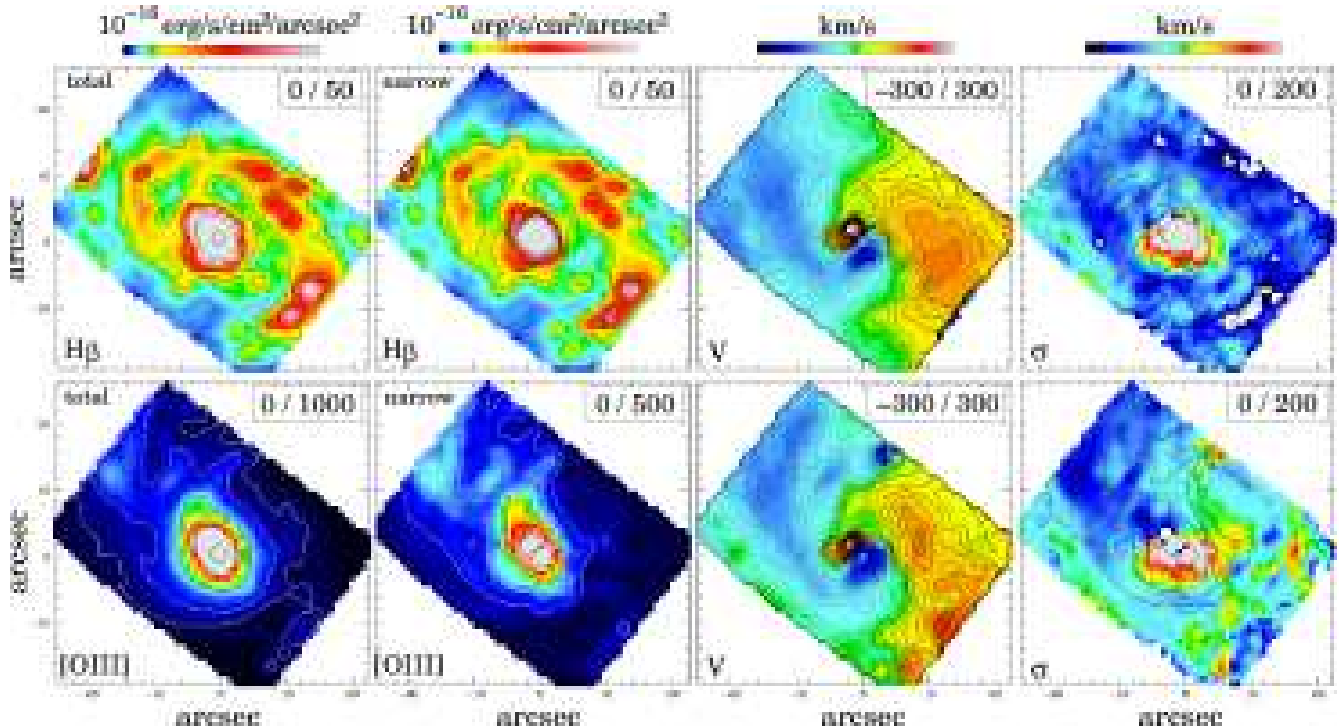, width=\hdsize}
\caption{\Sauron\ maps of the gas distribution and kinematics for NGC~1068.
Top row (from left to right): distribution of the total, and narrow
components, and velocity and velocity dispersion for the \Hb\ narrow
component. Bottom row: same as top row, but for the \OIII\ component. Boxes in the top right corner of each panel show the ranges covered by the color bars.}
\label{fig:gas}
\end{figure*}

\subsection{Gas distribution and kinematics}
\label{sec:gas}

The maps of the distribution and kinematics of the \Hb\ and \OIII\ components, 
as reconstructed from the results of multiple Gaussian
fits to the data, are displayed in Fig.~\ref{fig:gas}.

\Hb\ emission from the narrow component is ubiquitous in our field of view but
the brightest region corresponds to the inner 3\arcsec\ of the galaxy. Away from
these nuclear regions, the narrow-component \Hb\ emission is dominated by the
contribution from the spiral arms. The distribution of \Hb\ shows the known
northern and southern spiral arms although they seem almost connected at this
resolution. Comparison between the total (all kinematic
components included) and narrow-component-only \Hb\ maps (top left panels
of Fig.~\ref{fig:gas}) show little difference between the two maps
outside from the nuclear regions, outlining the fact that the narrow
component dominates the \Hb\ emission away from the nucleus. 

As for \Hb, emission from the inner 3\arcsec\ of the galaxy also dominates the
narrow-component \OIII\ map. However, away from the nucleus the distribution of
the \OIII\ emission differs significantly from the distribution of the \Hb\ one.
The distribution of the \OIII\ emission is very asymmetric, thus 
found predominantly North-East of the nucleus, and does not trace the spiral arms. 
Instead, it traces the northern ionization cone (see e.g. Pogge 1988).

\begin{figure}
\epsfig{file=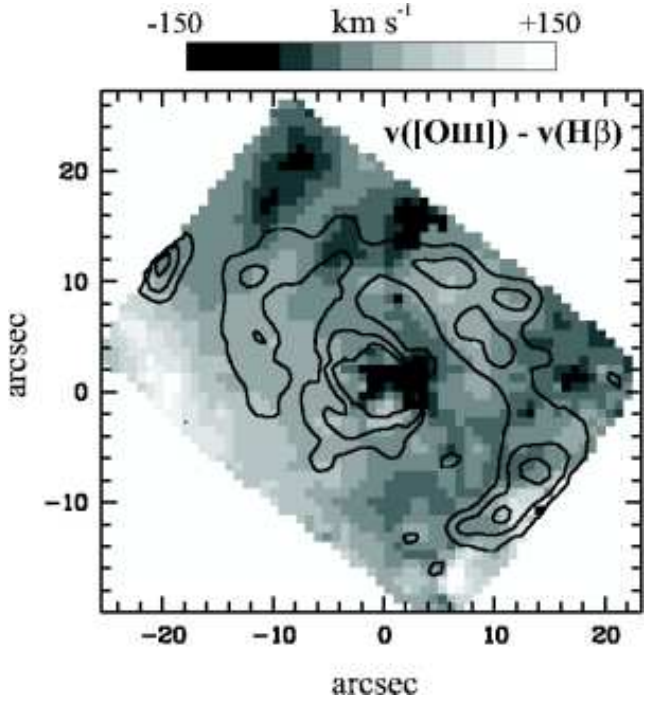, width=\columnwidth}
\caption{
Reconstructed map of the difference between the centroid velocities of the \OIII\ and \Hb\ lines of the narrow component, with contours of the \Hb\ narrow-component map overimosed. Contours: 10, 20 and 40 \ergscm. 
\label{FigureDeltaV} }
\end{figure}

Despite the differences between the distribution of \Hb\ and \OIII,
the overall morphologies of the \Hb\ and \OIII\ velocity fields (see
Fig.~\ref{fig:gas}) are very similar. They both display the prominent
"S" shaped structure, which is already present in the data of Cecil \etal 
(1990; see also Sect.~\ref{sec:HaCO}) and of
Garc\'{\i}a-Lorenzo et al. 1999 (their Fig.~16). 
However, if we take a closer look at these velocity maps, 
we see differences between the observed \OIII\ and \Hb\ velocities in some
regions. We have therefore built a map of v(\OIII) $-$ v(\Hb), which is
displayed in Fig.~\ref{FigureDeltaV}. Differences up to $\pm$150~\kms\ are
measured, with uncertainties of typically 40~\kms\ (3$\sigma$). Strong
positive values of observed v(\OIII) $-$ v(\Hb) are located in the South-East half of
our field of view, while strong negative values are found on the other half.
These qualitatively follow the perturbations observed in the stellar kinematics (see
Sect.~\ref{sec:skin}).
\begin{figure}
\epsfig{file=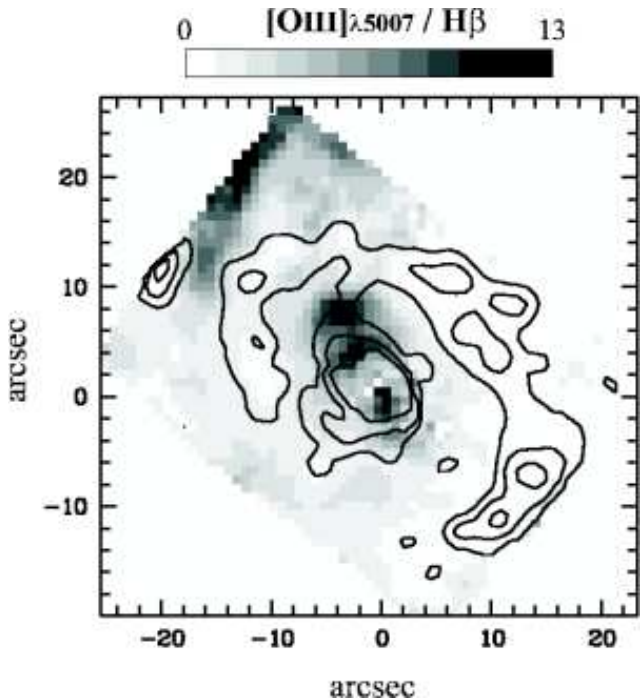, width=\columnwidth}
\caption{ Reconstructed map of the \OIIIwb\ / \Hb\ ratio, with contours of the \Hb\ narrow-component map overimosed. Contours: 10, 20 and 40 \ergscm. A clipping has been applied to the map, putting to 0 pixels where the relative accuracy of the ratio was worse than 20\% (only a few pixels south-west of the nucleus). 
\label{FigureOIIIHb} }
\end{figure}

We also show the velocity dispersion maps of \Hb\ and \OIII\ (right panels of
Fig.~\ref{fig:gas}). In both maps and if we exclude the nuclear regions, the
regions of bright emission display a dispersion lower than their
lower-surface-brightness surroundings. We checked that this was not a bias
introduced by differences in signal-to-noise ratio between bright and weak
regions\footnote{There is a systematic trend to obtain larger dispersion values as a
result from the fit when the signal to noise ratio in a spectrum decreases.}.
The signal to noise ratio in our \Hb\ and \OIII\ observations is $> 20$ over
most of our field of view and our Monte-Carlo simulations (see
Sect.~\ref{SectionELfitting}), show that, in this case, our relative
uncertainties on the measured dispersion values are $< 10$~\% (3$\sigma$). We observe
differences of the order of 0.5~\AA\ for dispersion values of typically 2.5~\AA\ and this
trend is therefore real. 

Last, we have built the map of the \OIIIwb\ / \Hb\ ratio (see
Fig.~\ref{FigureOIIIHb}). The range of variation of this ratio is extremely
large with \Hb\ dominating in the arms (ratio below unity) and \OIII\ dominating
in the ionization cone (ratio reaching values up to 13-14). 
The peak of high \OIII\ / \Hb\ ratio around 
a PA of 30\degr\ corresponds to the northern ionization cone. Note that the presence of
the well-defined peak suggests that, within the cone, the excitation of the gas
decreases from its center to its edges. This result must however be taken with
caution as the regions responsible for this peak are located at the very edge of our
field of view. 

%
There does not seem to be a systematic association between regions which exhibit
large velocity differences between \OIII\ and \Hb\ (Fig.~\ref{FigureDeltaV}), and those with high
\OIII\ / \Hb\ ratio (Fig.~\ref{FigureOIIIHb}). Outside a radius of 5\arcsec, 
we observe large velocity differences up to 100~\kms\ 
and rather normal \OIII\ over \Hb\ line ratios. This suggests that the
observed kinematic disturbance is not directly linked with the AGN activity. 

\subsection{Comparison with the \Ha\ and CO maps}
\label{sec:HaCO}

\begin{figure*}
\centering
\epsfig{file=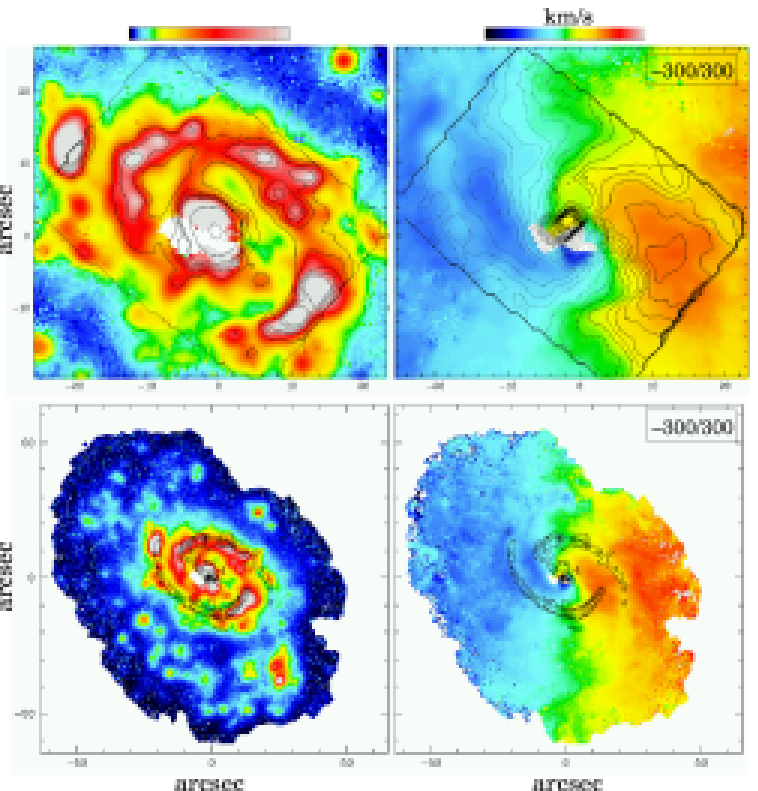, width=\hdsize}
\caption{Top panels: comparison between the \Sauron\ H$\beta$ flux (contours, left panel) 
and velocity (contours, right panel) and the corresponding H$\alpha$ maps obtained by 
Bland-Hawthorn \etal (1991) with a HIFI Fabry-Perot. The velocity step used for
the \Sauron\ isovelocities is 25~\kms. Bottom panels: comparison of the \Ha\ maps
with the CO distribution (contours: Schinnerer \etal 2000). Boxes in the top right corner of each panel show the ranges covered by the color bars.}
\label{fig:compHa}
\end{figure*}

The agreement between the \Sauron\ H$\beta$ flux and velocity
maps with the corresponding \Ha\ maps,
as shown in Fig.~\ref{fig:compHa}, is remarkable considering the very
different instrumental setup, and the fact that dust extinction is present
(Bruhweiler \etal 2001). Apart from the central few arcseconds, where 
scattering effects are important and AGN related emission is present,
extinction significantly affects \Hb\ 
(as probed by a ratio of \Ha\ over \Hb\ well over 3; Bruhweiler \etal 2001) in the South-West
part of the arm (SW clump at a radius of about 14\arcsec; Bruhweiler \etal 2001).
Both the \Ha\ and \Hb\ velocity field exhibit a very strong spiral-like
perturbation with the zero velocity curve displaying a very wavy shape.
The \Ha\ map is analysed further in Sect.~\ref{sec:ring}.

A comparison between the maps of the $^{12}$CO(1-0) line flux and 
the corresponding \Ha\ emission is shown in Fig.~\ref{fig:compHa}.
They both roughly follow the two-arm spiral structure with a diameter of
about 40\arcsec\ (often mentioned as the circumnuclear ring). 
There are however some very significant differences which we
emphasize here. Firstly, the \Ha\ emission is much more asymmetric with respect
to the centre with a brighter northern arm. Secondly the CO spiral is clearly offset 
from the \Ha\ arms: it is on the inner side of the SW \Ha\ clump, 
but seems to lie outside the \Ha\ arm in the North-West. A similar comparison
at higher spatial resolution using the HST/WFPC2 images (Bruhweiler \etal 2001) 
shows that the spiral arms seen in CO and \Ha\ are in fact offset from each
others (the northern and southern CO arms being outside and inside the corresponding \Ha\ arms, 
respectively). This is commonly observed in spiral barred galaxies (Sheth \etal
   2002). In the case of NGC~1068, the offset could be partly explained by the fact that most 
of the \Ha\ emitting regions are associated with starbursting HII regions. Davies \etal (1998) 
thus showed that most of the young stars in the ring formed in compact clusters in a relatively recent 
short burst within the last 30 Myr, about ten dynamical timescales
at the radius of the ring. But extinction also plays a role here since $A_V$ of $\sim 1-3$~mag 
have been measured for these clusters (Davies, Sugai \& Ward 1998). We therefore do not
expect a perfect coincidence between the CO and \Ha\ emission line gas
(according to Davies \etal 1998, the ionization cone cannot have a
significant effect on the star formation). Since the near side of the disc is South of the nucleus,
we cannot discard the possibility that we only clearly detect HII regions closer
to the edge and in front of the molecular arms, which would explain the relative
locations of the ionized and molecular arms. We however do
not have a good tracer of the currently ongoing star formation compared to the slightly older HII
regions, and it is therefore not possible to distinguish between the scenarios of 
dissociation of molecular gas due to hot stars, or segregation of CO and HII regions (as seen
in the spiral arms of M83, M100, or M51, see e.g. Rand \etal 1999) due to a
spiral density wave, or even some other unknown effect: 
observations of mid-infrared emission lines are required to test this.


\subsection{Harmonic Analysis of the \Ha\ Velocity Field}
\label{sec:ring}
\begin{figure*}
\centering
\epsfig{file=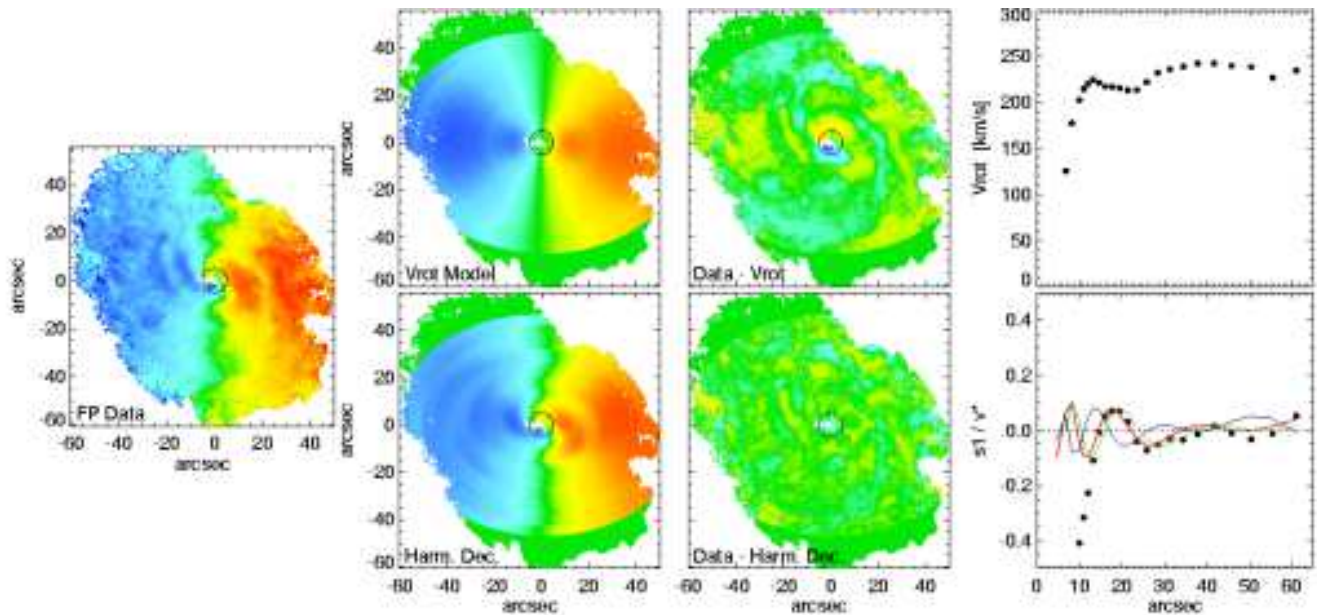, width=18cm}
\caption{Left panel: observed \Ha\ velocity field. Top row, from left to right: best fit circular
velocity component alone, corresponding residual map (Data - \Vrot\ model) and the derived
rotation curve assuming an inclination of $40\deg$, PA$=87\deg$, and \Vsys$=1140$~\kms.
Bottom row, from left to right: harmonic reconstruction including the rotational velocity component,
corresponding residual map (Data - Harmonic Decomposition Fit) and the derived radial
velocity component normalized with $v^* = V_\mathrm{rot} \sin(i)$ as a function of
galactocentric radius. The solid curves in the bottom right panel show the radial velocity
component according to the analytic spiral model as described in Sect.~\ref{sec:ring}. The best matching
spiral model with CR at 31\arcsec\ corresponding to the spiral pattern $\Omega_p = 109 \pm 5$
 \kmskpc\ is shown in red. Radial velocities for spirals with pattern speeds of 99 and 119 \kmskpc\
are shown in green and blue colours respectively.}
\label{fig:tiltedrings}
\end{figure*}
Our \Sauron\ \Hb\ and \OIII\ velocity fields, as well 
as the \Ha\ one, exhibit complex kinematic features such as 
``S''-shape or wobbles in the zero velocity curve. We aim to extract the 
effect of the prominent NIR bar and that of the spiral arms from the velocity field of
NGC~1068. Given that the \Ha\ field covers a significantly larger area than the 
\Sauron\ field (at the expense of a shorter spectral coverage), we 
will therefore apply our analysis method on the former.
Assuming that circular motion is the dominant feature, and that there 
is no strong warp in our field, we carry out an analysis 
based on the harmonic decomposition of the line-of-sight velocity \Vlos\  
(Schoenmakers, Franx \& de Zeeuw 1997; see also Franx, van Gorkom \& de Zeeuw
 1994). This formalism implies expanding 
the \Vlos\ field into harmonic series, where the first terms 
($\cos\theta$ and $\sin\theta$) are the rotational and radial velocity 
components, and the higher harmonic terms ($\cos m\theta$ and $\sin m\theta$) 
provide information about perturbations on the gravitational potential 
(Wong, Blitz \& Bosma 2004; Fathi 2004).
The first, second, and third harmonic components, are sufficient for studying 
specific elements of the perturbations on the underlying 
potential. A perturbation of order $m$ creates $m-1$ and $m+1$ 
line-of-sight velocity terms (eg. Canzian 1993 and Schoenmakers, Franx \& de Zeeuw 1997), 
i.e. third harmonic terms contain information about an 
$m=2$ bar or a two-arm spiral perturbation. 
 
To obtain the kinematic PA, the systemic velocity, and circular
velocity contribution to the \Ha\ \Vlos\ as a function of galactocentric
radius, we apply a tilted-ring method similar to that by
Begeman (1987). We first assume an inclination of $40\deg$,
then for each galactocentric radius, we find the best fit PA and
systemic velocity. These parameters did not show a significant
variation throughout the field: the PA variation is found to be
less than $10\deg$ and the \Vsys\ varies less than 20\kms.
Following the standard procedure, we then fix the PA and \Vsys\ to
the average values. This step yields PA$=87\deg$ and
\Vsys$=1140$~\kms, which are consistent with the values mentioned
in the rest of the present paper. Keeping these parameters fixed, we now
fit the rotational component, and derive the rotation curve presented in
Fig.~\ref{fig:tiltedrings}. Reconstruction of the circular velocity
component and subtraction from the observed velocity field yields
the residual non-circular velocities as presented in Fig.~\ref{fig:tiltedrings}. 

We explore the non-circular velocity components by fitting the 
residual map with the higher order harmonic terms up to and including order 3.
The second harmonic terms are mainly consistent with zero. 
The first and third harmonic terms show a behaviour similar to that 
of an analytically derived logarithmic two-arm spiral perturbation 
(Wong et al. 2004). We therefore construct a library of velocity fields
perturbed by two armed spirals with different spiral structure characteristics. 
The library of models include a range of pitch angles, perturbation amplitudes,
and spiral arm sizes. We compare the harmonic terms and non-circular velocity features
with those of the models to find a model which best 
resembles the observed features. We find that the spiral model with a pattern
speed $\Omega \sim 109 \pm 5$~\kmskpc, a pitch 
angle of $15\deg$, a spiral amplitude of $0.17$, and with its corotation radius 
at 31\arcsec\ provides the best fit to the observed \Ha\ velocity field. 
In the region of the inner near-infrared bar, there is however significant 
disagreement between this model and the observed maps, 
the spiral model alone being obviously inadequate for describing the observed features.

\subsection{Application of the Tremaine-Weinberg method}
\label{sec:tw}

\begin{figure}
\centering
\epsfig{file=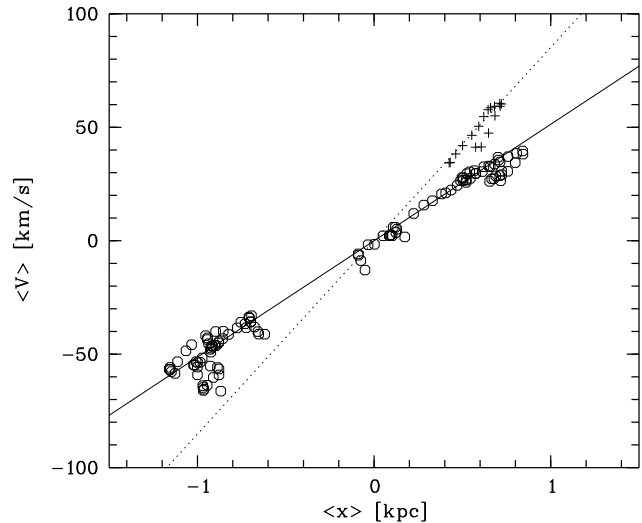, width=\columnwidth}
\caption{Tremaine Weinberg method applied on the \Ha\ Fabry Perot maps of  Bland-Hawthorn \etal (1991).
The averaged, density weighted, velocity $\langle V \rangle$ is plotted against the averaged,
density weighted coordinates $\langle X \rangle$, along slits parallel to the line of nodes
assuming PA$=90\degr$. The best fit slopes for the outer points (open symbols, $R >
20\arcsec$) and inner points (crosses, $8\arcsec< R < 15\arcsec$) are shown as a solid and dotted
line, respectively. }
\label{fig:omega}
\end{figure}

Rand \& Wallin (2004; RW04 hereafter) recently argued for the presence 
of two different pattern speeds for the NIR inner bar and the spiral 
arms which surrounds it in NGC~1068. They quote a value
of $\Omega_p \sim 135 \pm 42$~\kmskpc\ for the inner bar and a lower value of
$\Omega^S_p \sim 72 \pm 4$~\kmskpc\ for the outer spiral arms via the Tremaine-Weinberg
method (Tremaine \& Weinberg 1984) applied to the CO molecular line data. This estimate for 
the pattern speed of the inner bar should be taken as a lower limit, considering
the presence of a second outer tumbling component. The validity of using this technique
on a gaseous component was tested by RW04 with
Nbody $+$ SPH simulations. A similar test was also recently performed by
Hernandez, Wozniak, Carignan \etal (2005) who emphasized the need to avoid
regions of shocks. We attempted to estimate the pattern speed of the spiral and
bar structures in NGC~1068 via the same technique but on the large-scale \Ha\ maps 
of Bland-Hawthorn \etal (1991): these clearly have a higher resolution 
(and thus provide more apertures) but a more clumpy distribution than the CO maps.
We confirm the trend found by RW04 and clearly detect two different 
patterns, for points inside or outside 20\arcsec.
The scatter of the $\langle V \rangle$ versus $\langle X \rangle$ diagram (see Fig.~\ref{fig:omega})
is small for the outer region with a well-defined slope
of $80 \pm 2$~\kmskpc\ for a PA of 90\degr, only slightly higher than the
value found by RW04. However, this slope strongly varies
from 56 to 104~\kmskpc\ when we assume a PA varying from 84 to 95\degr,
a dependency already emphasized by RW04 and Debattista (2003; see also Debattista \& Williams 2004). 
The pattern speed of the inner region is rather badly constrained by our data,
an expected result considering the presence of star forming regions and the
influence of the ionization cone. Using the odd parts of $\langle V \rangle$ and 
$\langle X \rangle$ data points to minimize the scatter 
(Tremaine \& Weinberg 1984), we find a lower limit value of $133 \pm 12$~\kmskpc\ 
for PA$=90\degr$ (consistent with the value of RW04), and from 93 up to 178~\kmskpc\ for PAs 
again varying from 84 to 95\degr. 

\section{N-body modelling}
\label{sec:dynmodel}

We now turn to the dynamical modelling of the central region of NGC~1068
using N-body and SPH simulations. This will be compared in Sect.~\ref{sec:resultsnbody}
with the \Sauron\ maps of the gas and stellar kinematics, as well as with the
available HI, H$\alpha$ and CO data (see Sect.\ref{sec:otherdata}).

\subsection{Methods}

We used PMSPH, the N-body code initially developed in Geneva (Pfenniger \& Friedli 1993). 
The gravitational forces are computed with a particle--mesh method using a
3D polar grid with $(N_R, N_\phi, N_Z)=(40,32,64)$ active cells. 
The hydrodynamics equations are solved
using the SPH technique (Friedli \& Benz 1993). Since the radial
spacing of the cells is logarithmic, the cell size increases from
15~pc at the centre to 383~pc at a radius of 2~kpc. 20 cells
(i.e. half the number of radial cells) are used to described
the region enclosed by \Sauron\ field. The total grid has a radius of
100~kpc. All runs were performed using a total of 978\,572 particles
for the stellar part and 50000 for the gaseous one. The initial
conditions (positions, velocities, velocity dispersions) were built
from a Monte Carlo realization of a five-component axisymmetric
model, as described below, and the total running times for each
simulation varied from 500 to 2000 Myr.

\subsection{Mass model and initial conditions}

We constrained the initial axisymmetric mass model of NGC~1068 by using the available
photometry and velocity curve as constraints.
We performed this in two steps. We first constrained the projected luminosity
distribution of the galaxy by combining a high resolution
near-infrared K band image (see Sect.~\ref{sec:otherdata}) with
the 2MASS K band image, the DSS I band image and a deep wide-field B band
image (see Sect.~\ref{sec:photom}). The near-infrared images clearly reveal the
bar within the central 20\arcsec, and the optical images allowed us to 
estimate the contribution of the outer oval and disk structures (see Sect.~\ref{sec:otherdata}).
We assumed the presence of a central spherical bulge and fitted a projected Plummer
sphere to the high resolution NIR K band image. 
This component was subtracted from all images,
which were then simply deprojected by assuming an inclination angle of $i=40\deg$, and a
two-dimensional distribution: this corresponds to a spatial scale factor of $\sim 1.3$ perpendicularly to
the line of nodes.
The deprojected images were found to be reasonably approximated by the sum of 
four individual Miyamoto-Nagai components, each of them probing a different scale (see Table~\ref{tab:MN}).

\begin{figure}
\centering
\epsfig{file=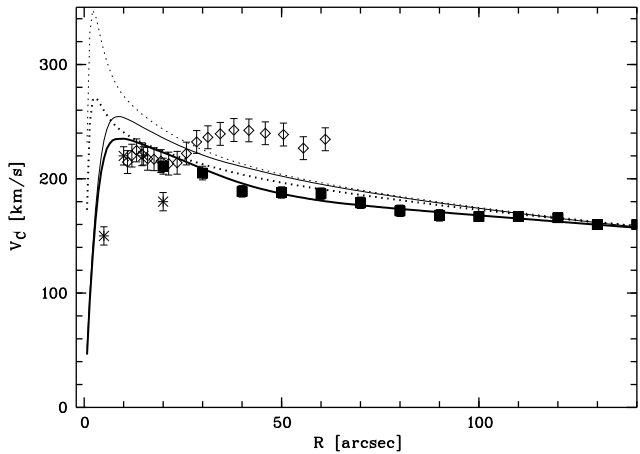, width=\columnwidth}
\caption{Circular velocity profiles of the Nbody$+$SPH simulations at
$t=0$ (dotted lines) and at the end of the runs (solid lines) compared
with observed velocity profiles. $t = 850$ for model A (thick lines)
and $t = 1440$ for model B (thin lines). Symbols show the deprojected
HI (filled squares), \Ha\ (diamonds; from this paper) and CO
(stars) velocity profiles (assuming an inclination of $i=40\deg$).}
\label{fig:Vc}
\end{figure}

We normalize each component, i.e. the four Miyamoto-Nagai, the Plummer sphere and one
additional Miyamoto-Nagai component for the gas, so that the resulting circular velocity
curve fits the observed HI velocity profile. This requires the overall mass-to-light ratio to increase 
by a factor of 10 between the central part and the outer most point of the HI curve
($R = 180\arcsec$). Since we know that the simulation will rapidly depart from its initial
conditions, this procedure is only meant to produce a three dimensional
mass distribution which reproduces the radial mass gradient,
the central bulge concentration and the large scale velocity curve reasonably
well. We consider the contribution of a presumed central black hole to be 
part of the Plummer sphere. The initial radial profiles of the circular velocity
are presented in Fig.~\ref{fig:Vc}.

\begin{figure}
\centering
\epsfig{file=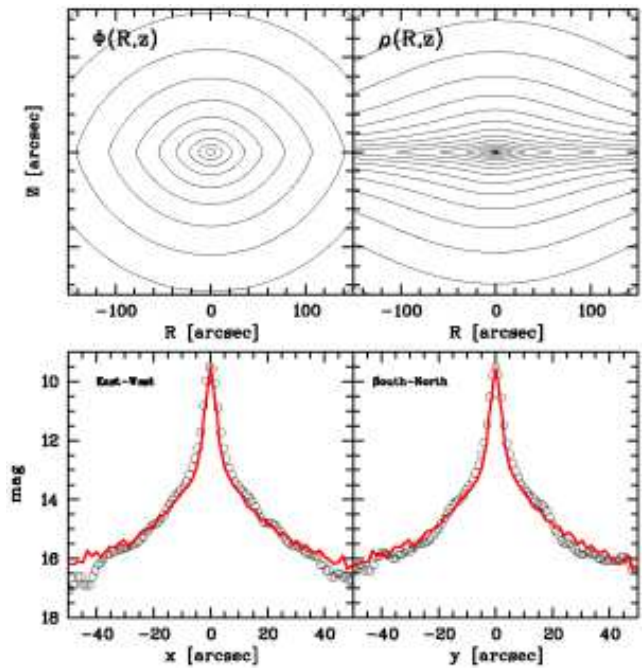, width=\columnwidth}
\caption{Initial conditions for model~B. Top panels: potential (left) and density (right)
   isocontours in the meridional plane $(R, z)$.  Contour steps are 0.1 and 0.5 in log respectively.
   Bottom panels: (projected) surface brightness profiles along the East-West
   (left) and South-North (right) direction extracted from the K band 2MASS
   images (open circles) and model~B after projection using an inclination of
   40\degr (solid lines).}
\label{fig:potrho}
\end{figure}

We performed a number of runs (36), where we mainly varied the concentration of
the different components, keeping the total mass (including the gas) roughly constant
(see below). We chose to ignore the observed \Ha\ velocity profile as a constrain
for the initial conditions of our models as it exhibit strong non-circular motions (e.g., see Sect.~\ref{sec:ring}).
We present only two models here (A \& B), as they are typical of the runs performed and
their initial circular velocity profiles roughly bracket the observed HI profile of
NGC~1068 (Fig.~\ref{fig:Vc}).  
Both models have initial conditions which are consistent with a
deprojected axisymmetric version of the available K band images. This is
illustrated for model~B in Fig.~\ref{fig:potrho}, in which we show the projected mass profiles
compared to 2MASS K band surface brightness profiles.
The corresponding gravitational potential and density distributions in the meridional plan 
are presented in the same figure. 

\begin{table}
\caption{Parameters of the initial conditions used in the N-body $+$ SPH simulations.
M-N and P stand for Miyamoto-Nagai and Plummer components respectively.}
\label{tab:init}
\begin{tabular}{lllcccc}
\hline
Component & Type & \# & Mass & a & b \\
 &  &  & [10$^9$ M$_{\odot}] $ & [kpc] & [kpc] \\
\hline
\hline
\multicolumn{6}{c}{Model A}\\
Stars & M-N & 1 & 4.4 & 0.255 & 0.035 \\
Stars & M-N & 2 & 18.7 & 1.0 & 0.1 \\
Stars & M-N & 3 & 46.2 & 5.0  & 0.2\\
Stars & P & 4 & 3.3  & -  & 0.1\\
Gas & M-N & - & 3.4 & 6.0 & 0.2\\
\hline
\hline
\multicolumn{6}{c}{Model B}\\
Stars & M-N & 1 & 2.4 & 0.155 & 0.035 \\
Stars & M-N & 2 & 18.7 & 0.8 & 0.2 \\
Stars & M-N & 3 & 46.2 & 4.6  & 0.4\\
Stars & P & 4 & 3.3 & -  & 0.1\\
Gas & M-N & - & 3.4 & 6.0 & 0.2\\
\hline
\hline
\end{tabular}
\label{tab:MN}
\end{table}

\subsection{Models A and B}
\label{sec:resultsnbody}

A bar starts forming around 300 Myr, but becomes prominent only after 900~Myr
for Model A. The bar forms much later on in Model B, this being mainly due to 
its significantly higher mass concentration. The overall evolution of Model B is
much slower for the same reason. In simulations with even higher mass concentrations
the disk was found to be stable against the formation of a bar. 
It is important to note
that the central Plummer sphere has a characteristic scale (100~pc), 6
times the central resolution grid point (15~pc).
After the bar formation, the gas wraps around in a low constrast multi-arm 
spiral structure for a few 100 Myr. 
The gas then reacts quickly to the formation of the $m=2$
perturbation and is radially redistributed. Part of the gaseous component flows
towards the central 500~pc, exhibiting a time varying structure. 
We derived a snapshot of both simulations after the bar is well formed, 
the time of each snapshot being chosen ($t=850$ for model A, and 1440~Myr for Model B)
such as to resemble the overall structure of the gaseous component in NGC~1068.
We optimized the viewing angles for the bar to lie at a PA of $44.5\deg$ as in NGC~1068,
and the zero velocity curve to be as close as possible to the observed one.
This requires (for a fixed inclination of $40\deg$) that the line of nodes is 
at a PA of $\sim 83\deg$ and $84\deg$ for models A and B respectively. 
We have tested the dependence of this value on the time of the snapshot as
well as on varying initial conditions: the resulting PA of the line 
of nodes varies then between 80 and $90\deg$. 

\subsection{Comparison with the observed stellar kinematics}
\label{sec:nbodycomp}

The simulations performed for this study are intended as a proof of
concept, and not as perfect fits to the data. The kinematics are
luminosity-weighted and we did not include star formation or
dust extinction in our simulations, although it is clear that young stars 
and dust significantly affect the surface brightness at the wavelengths 
of the \Sauron\ observations. Furthermore, the bars formed in these simulations do not
perfectly match the well-known near-infrared bar, and we could not reproduce
the large-scale (North-South) oval in the photometry (the suggested signature of a large-scale bar). 

Model~A fits the HI inner velocity profile better than model~B (both models
having similar circular velocities outside 100\arcsec), 
but significantly fails to reproduce the first two stellar velocity moments (velocity and velocity dispersion),
with values significantly below the observed data. We will therefore focus on the 
results from model~B in the rest of the paper.
The bar in model~B has a pattern speed of $\sim 100 $~\kmskpc, which leads to radii
for the Inner Inner and Outer Inner Lindblad resonances (IILR, OILR), 
4:1 ultra harmonic (UHR), Corotation (CR), and Outer Lindblad resonances (OLR)
of 5\farcs2, 10\farcs5, 22\farcs6, 32\farcs2, 50\farcs2, respectively
(0.36, 0.73, 1.58, 2.25, 3.50 kpc respectively). This
implies that the NIR bar ends well inside its CR, closer to its UHR.
\begin{figure*}
\centering
\epsfig{file=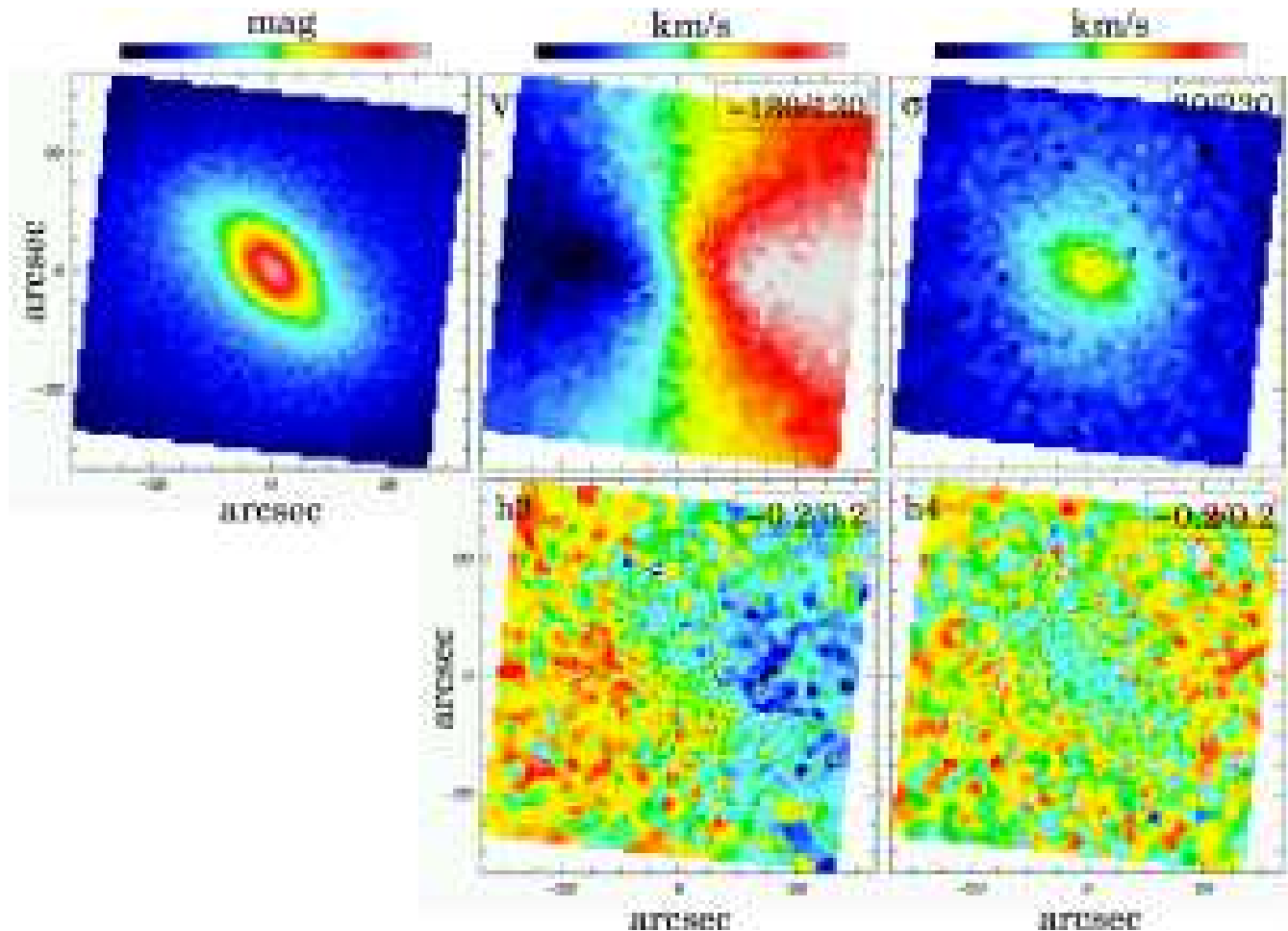, width=\hdsize}
\caption{Luminosity distribution (top left panel, assuming a constant mass-to-light ratio),
   and observed stellar kinematics from the model B at $t=1440$
(using an inclination of $i=40\deg$). The panels correspond to the ones in Fig.~\ref{fig:starkinmaps}.
Boxes in the top right corner of the panels show the ranges covered by the color bars.}
\label{fig:stelkin_nbody}
\end{figure*}

\begin{figure}
\epsfig{file=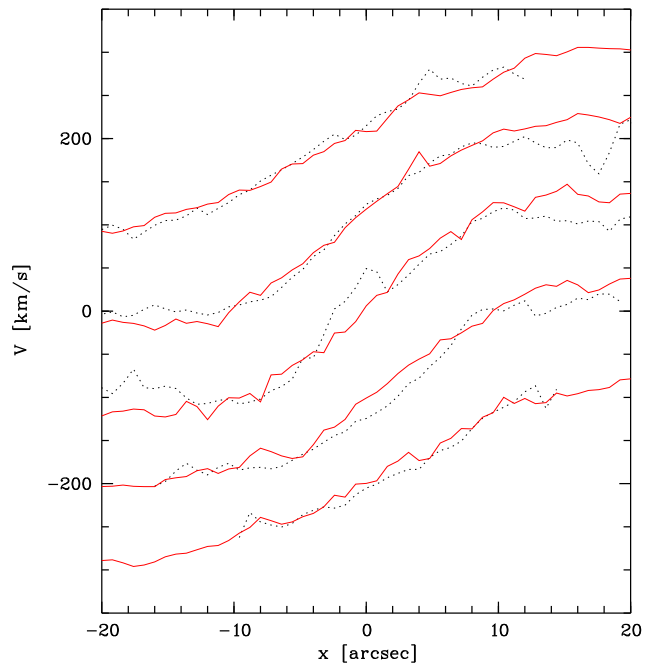, width=\columnwidth}
\caption{Comparison between stellar kinematic profiles extracted from the
two-dimensional \Sauron\ velocity field (solid lines) and from model B at $t=1440$~Myr
(dashed lines).
Four cuts along PA$=90\deg$ are shown, with offsets, from top to bottom,
of 10, 5, 0, -5, and -10 arcseconds respectively.}
\label{fig:sauronmod}
\end{figure}

A comparison of the stellar kinematics of model B
within the \Sauron\ field of view is
presented in Fig.~\ref{fig:stelkin_nbody}: it looks qualitatively
similar to the observed \Sauron\ maps. The twist in the 
stellar velocity field is less pronounced in the model, which 
nevertheless shows the clear signature of non-axisymmetry.
A more quantative comparison is shown in Fig.~\ref{fig:sauronmod},
where a few cuts at PA$=90\deg$ of the stellar
velocity fields are presented against the \Sauron\ data.
The overall agreement is quite good, with the exception
of the outer parts of the central profile in which there is
a discrepancy of about 20~\kms: this is where we detected
perturbations in the stellar kinematics (Sect.~\ref{sec:skin}). As mentioned already,
they follow the velocity differences between the \Hb\ and \OIII\ emission lines
(Sect.~\ref{sec:gas}). The discrepancy between the model and observed
stellar kinematics therefore mostly reflects the presence of streaming motions
within the outer spiral arms, where star formation is on-going.
Model~B additionally reproduces the morphology of the observed $h_3$ field, 
with the zero $h_3$ curve almost following the major-axis of the bar, as in 
the \Sauron\ map. There is only a slight central depression in the $h_4$ map of
model~B, less pronounced than in the data. The model does also not reproduce the
ring-like enhancement of $h_4$ observed both in the \Sauron\ and long-slit data.

\begin{figure}
\epsfig{file=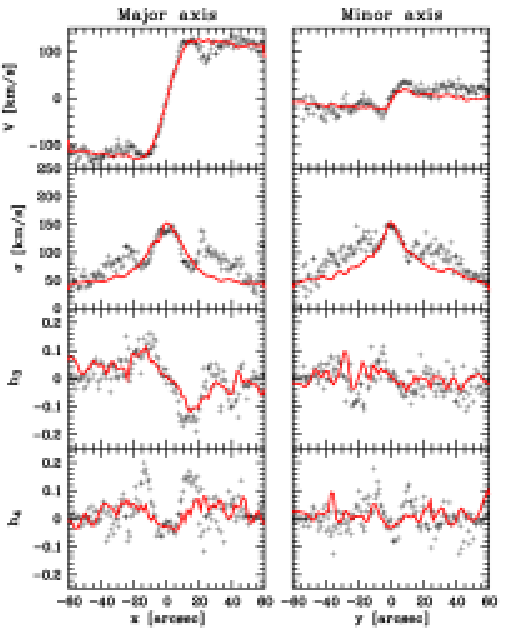, width=\columnwidth}
\caption{Comparison between stellar kinematic profiles extracted from the
Sh+03 (crosses) and from model B at $t=1440$~Myr (solid
lines). From top to bottom: mean velocity, velocity dispersion, $h_3$ and $h_4$. The model kinematics
have been averaged over a slit width of 3\arcsec.}
\label{fig:shapiromod}
\end{figure}

We can now compare the obtained stellar kinematics with the more extended
long-slit kinematics obtained from the data of Sh+03: a
comparison is presented in Fig.~\ref{fig:shapiromod}. The agreement is again
very reasonable, including the higher order Gauss-Hermite moments. 
The kinematic profiles of the simulations do not however exhibit the characteristic
dispersion peaks at a radius of $\sim 22\arcsec$ on either sides of the center along the major-axis.
These peaks correspond to a transition region at the end of the bar, where the
slope of the velocity gradient also changes, and are close to the
UHR in model~B. The simulations also lack the central plateau or depression
in the dispersion profile observed in the central 5\arcsec\ of NGC~1068.

\section{Discussion}
\label{sec:discuss}

\subsection{The location of resonances}
\label{sec:resonances}
Both the molecular and ionized gas kinematics exhibit strong departures from circular motion
(see e.g., Fig.~\ref{fig:tiltedrings}), and
the presence of the near-infrared bar in the inner 1.5~kpc is undisputable.
The \Ha\ (Fig.~\ref{fig:compHa}) and CO velocity fields
(S+00) provide us with an important clue regarding
the kinematics of the spiral arms: the isovelocities clearly bend
outwards which implies an inward streaming motion (e.g. see the spiral arm
velocity contours at about 10\arcsec\ South, 5\arcsec\ West). A similar argument
was initially emphasized by Yuan \& Kuo (1998), although they reached the opposite
conclusion as they relied at the time on the CO data of Helfer \& Blitz
(1995) which lacked spatial resolution and seemed to show an inward 
bending\footnote{Yuan \& Kuo (1998) also noted that they "would not be surprised"
to get a different result when a more reliable rotation curve becomes
available.}. The inward streaming therefore
requires the gas to be located inside the corotation (CR) or outside 
the outer Lindblad resonance (OLR) of the corresponding density wave,
a conclusion also reached by S+00 and RW04 (and opposite to the one
of Yuan \& Kuo 1998, but see above).

The presence of a large-scale tumbling component with a radius of
about 9~kpc is debatable, although a very significant elongation
is clearly observed in the photometry at a PA near the minor-axis, 
inconsistent with an axisymmetric outer disk. 
S+00 first estimated the pattern speed of the outer 
oval\footnote{S+00 quotes a value of $\Omega_p = 20$~\kmskpc\  in their paper,
although, as mentioned by RW04, they actually used a value of 35~\kmskpc.} 
to be $\Omega_p \sim 35$~\kmskpc.
Assuming the ILR of the primary bar corresponds to the CR of the inner (secondary) bar, 
they determine its pattern speed to be $\Omega_s \sim 140$~\kmskpc.
This would imply that the gaseous spiral arms lie between the CR and 
ILR of the primary bar. These arms would however lie outside the 
CR of the secondary bar, and the assumed value for $\Omega_s$ would
make the near-infrared bar extend as far as its CR. Recent simulations
of double bars (Rautiainen, Salo \& Laurikaine 2004) seem indeed to favor
CR/ILR coupling (but see Moellenhoff \etal 1995), but
with short secondary bars, ending inside the 4:1 Ultra Harmonic Resonance (UHR).
This requires the pattern speeds for both the primary and secondary bar to be
lower.

Assuming the PA of the line of nodes is between 84 and 90\degr (Sect.~\ref{sec:nbodycomp}),
a lower limit for the pattern speed for the inner bar is estimated to be between 93 and 133~\kmskpc
(Tremaine-Weinberg method on the \Ha\ data, see Sect.~\ref{sec:tw}), 
while the bar that forms in Model~B has a pattern speed of 100~\kmskpc (Sect.~\ref{sec:nbodycomp}).
These values would be consistent with the inner bar ending inside its UHR as discussed
above. If we believe the outer oval corresponds to a third tumbling structure (S+00), 
a CR/ILR coupling between the two bars would then imply a primary bar with 
$\Omega_p \sim 25$~\kmskpc, lower than but still consistent with the value advocated by S+00.

As mentioned above, 
the N~body simulations conducted in this paper lead to a PA for the lines of nodes in the range
83--90\degr\ with a preferred value of 84\degr. 
This is consistent with the photometric PA of the outer disk, and with the output of the
tilted ring gas modelling (Sect.~\ref{sec:ring}). It is not consistent
with the HI kinematic axis determined by Brinks (1997). Brinks however noticed a
gradual change in the position angle of the HI kinematic major axis, which he
interpreted as the sign of a mild warping. The HI map exhibits a weak
spiral arm or ring-like structure in the radius range 130--180\arcsec\, 
the density decreasing then inwards (down to the
inner HI ring present between 30\arcsec\ and 80\arcsec; Brinks~1997). If the
outer oval is, as mentioned above, a weak bar tumbling at 25~\kmskpc, 
its Outer Lindblad Resonance would be located at a radius of $~140\arcsec$ (10~kpc),
within the outer HI structure. At the OLR, we expect periodic orbits 
to be elongated perpendicularly to the major-axis of the bar. The observed
radial variation in the HI kinematics could therefore be partly explained
by the resulting non-circular motions. This should however be 
confirmed by a detailed kinematic HI study in the outer region of NGC~1068.

We finally turn to the derivation of the pattern speed associated with the
outer spiral arms. Our estimate using the Tremaine-Weinberg technique is
between 56 and 80~\kmskpc\ (for PAs between 84 and 90\degr), consistent
with the values given by RW04. This is however significantly lower than the
estimate obtained via the harmonic decomposition (Sect.~\ref{sec:ring}), which
provides a value close to the estimate of the pattern speed of the NIR bar.
If the outer spiral and the inner bar
share the same pattern speed, it would hint at the bar being the main driver.
But the outer spiral and the inner bar having different pattern speeds 
would not be unexpected as modes can couple non-linearly (Masset  \& Tagger 1997). 
Such a process is witnessed in some simulations
of double bars (see e.g., Rautiainen \etal 2004) which exhibit up to
four different $m=2$ modes. It would however not be wise, considering
the large uncertainties in the determination of the tumbling frequencies,
to speculate further if non-linear coupling of modes are present in NGC~1068.
The pattern speed of the outer spiral is in all cases much higher than 
the one of the presumed large-scale bar.

\subsection{Signatures of the bar}

Bureau \& Athanassoula (2005) drew attention to a number of the stellar kinematics signatures of 
a bar using pure N~body simulations: a double hump velocity profile, 
correlated stellar $V$ and $h_3$ profiles within the bar, 
secondary peaks for the dispersion $\sigma$ near the end of the bar, a flat central minimum
in the $h_4$ profile followed by a significant rise and a decline at larger
radii. These features were however obtained for rather high inclination ($i > 75\degr$), 
and for a galaxy with a very prominent outer stellar ring.
There are secondary peaks in the observed dispersion profiles of
NGC~1068 at a radius of 22\arcsec\ for a PA of 80\degr (see e.g., 
Fig.~\ref{fig:shapiromod}), and a flat minimum in $h_4$ which corresponds to the
flat central part in $\sigma$ (within a radius of 5\arcsec).
But, although there is a local minimum in $V$ corresponding 
to the secondary $\sigma$ maxima, the double hump nature of the velocity profile is not
very pronounced: the outer stellar velocity curve is nearly flat with a level 
comparable to the inner maximum of $\sim 120$~\kms\ at radii between 10 and 15\arcsec.
We furthermore observe $h_3$ profiles which are clearly anti-correlated with $V$
within the extent of the bar: this means that we detect a tail of low velocity
stars superimposed on the rapidly rotating ones along the line of nodes.

The high resolution K band image of NGC~1068 obtained by Peletier \etal (1999) 
shows the inner bar with rather pinched ends at a radius of $\sim 16\arcsec$.
These can be interpreted as the tips of the $x_1$ orbits sustaining the bar.
There is a clear plateau in the surface brightness profile around that radius,
which is also the witness of a radial (and possible vertical) redistribution
of the stellar component (Combes et al. 1990).
The fact that we do not reproduce the higher stellar dispersion 
(and lower stellar velocity) around a radius of 22\arcsec\ (PA$=80\degr$, 
Fig.~\ref{fig:shapiromod}) with our simulations  may be due
to the lack of star formation in our models. This region in NGC~1068
is the site of intense star formation, and corresponds to the UHR of model~B:
stars often form near the UHR (Schwarz 1984, Buta \& Combes 1996), 
and gathering new stars with orbits at the UHR (4:1 resonance) 
may be the cause of an increase in the stellar velocity dispersion.

\subsection{Driving gas towards the central 300~pc}

Star formation could also be the cause of the discrepancy 
between the rather flat (or depressed) central dispersion profile 
of NGC~1068 and that of model~B.
The inner bar is able to drive gas inwards, as illustrated
by some numerical simulations of double barred galaxies (see Model~III 
of Rautiainen \etal 2004) although this is certainly not a generic
result for double bars (e.g., Maciejewski \etal 2002).
If we assume that the NIR bar ends well inside its CR, 
it then implies the presence of an Inner Lindblad Resonance around 5\arcsec.
This would roughly coincide with the extent of the stellar velocity
decoupling and $\sigma$-drop, and support the generic scenario 
outlined by Emsellem \etal (2001) and Wozniak \etal (2003):
the inner NIR bar has driven gas within its Inner Lindblad Resonance, thus
forming a relatively cold stellar system within the central few hundred
parcsec. 

We are indeed witnessing radial gas inflow in the central kpc in
our numerical simulations. The inflow rate in model~B ranges from
$10^{-2}$ to a few M$_{\odot}$/yr within the region of the bar, with an average
of about $5\; 10^{-2}$~M$_{\odot}$/yr. This gaseous mass inflow increases significantly as the
bar gets stronger. The net mass increase driven by the bar 300~Myr after its
formation is $\sim 5\; 10^7$M$_{\odot}$ inside 5\arcsec\ (350~pc), which represents about 1\% of
the total enclosed mass. All these values should be taken as upper limits as e.g., 
the simulations do not include star formation. In order to make the link with
still smaller scales, high resolution two-dimensional mapping of the
stellar kinematics and populations in the inner few arcseconds would be
required - a considerable challenge given the active and disturbed nuclear
environment of NGC~1068.

\section{Conclusions}
\label{sec:conclude}

We have successfully obtained stellar and gaseous kinematics for NGC~1068 over a
large two-dimensional field covering the entire near-infrared inner (secondary) bar. 
Although our spectra exhibit numerous emission and absorption features, 
we have been able to disentangle
the two components, and derive the kinematics independently. Our \Sauron\ data
reveal a regular stellar velocity field that rules out an offset 
stellar system as claimed by Garcia-Lorenzo \etal (1998). The \Sauron\
two-dimensional stellar kinematics exhibit the signatures of the presence of the inner bar:
a twist in the velocity field, and an $h_3$ field elongated along the bar.
These features are also retrieved in a
qualitative comparison with N-body $+$ SPH simulations. 

We detect a kinematic decoupling of the central 350~pc (5\arcsec) in
a reanalysis of previously published long-slit data (Sh+03). We
first observe a flattening of the $h_3$ profile indicating a change of 
orbital structure within this region. 
We then reveal a flattening or a potential drop in the central part of the
stellar velocity dispersion profile.  
The kinematics of the ionized \Hb\ and \OIII\ gas both show strong
inwards streaming motions. Differences up to 100~\kms\ are observed 
in the ionised gas kinematics inferred from the \Hb\ and \OIII\ lines.
The CO and \Ha\ distributions 
are not coincident, a difference which can be explained by the effect of
the on-going star formation, and dust extinction. 

We confirm the presence of two different pattern speeds for the region inside
and outside of the inner bar (RW04), applying the Tremaine-Weinberg method
to the \Ha\ velocity field. The pattern speed of the inner bar is however
only weakly constrained. Note that the outer oval could correspond to a third
tumbling structure, as mentioned by S+00.

We then performed numerical simulations with initial axisymmetric conditions
approximating the photometry of NGC~1068. We focus on one model which has a
gravitational potential consistent with the HI velocity profile, and 
in which a bar forms. After projection with the relevant viewing angles,
the resulting stellar kinematics successfully reproduce a number of properties
observed in the \Sauron\ and long-slit kinematics, including the $h_3$
and $h_4$ profiles. There are however some discrepancies between the model and
the observations, which could be partly explained by the exclusion of star
formation and the lack of a primary large-scale bar in our simulations.

We finally briefly discuss the possibility of gas fueling within the inner bar. 
The numerical simulations suggest that the inner bar could drive a significant amount of gas 
down to a scale of $\sim 300$~pc (its ILR). This would be consistent with the
interpretation of the $\sigma$-drop (Emsellem \etal 2001) in NGC~1068 
being the result of central gas accretion followed by an episode of star formation 
(Emsellem \etal 2001; Wozniak \etal 2003). We should however wait for a
detailed view at the stellar kinematics and populations within the inner 300~pc
to conclude on this issue, e.g. with an integral-field spectrograph in the near-infrared 
such as SINFONI at the VLT. It would also be important to examine a sample of
carefully chosen galaxies, as to allow a statistical analysis and to assess the
relative contributions of the various physical mechanisms at play.

\section*{Acknowledgments}

We are grateful to Joss Bland-Hawthorn for his help
with digging in the Ringberg standards, and for supplying the beautiful
\Ha\ maps used in this paper. We thank Kristen Shapiro and Joris Gerssen for useful 
discussions and for allowing us to use their long-slit data.  
We would like to thank Reynier Peletier for providing us
with the high resolution K band image used for the modelling. We would like to
emphasize the support from the WHT staff, which allowed this work
to be conducted. We wish to warmly thank Michele Cappellari, 
Jesus Falcon-Barroso, Harald Kuntschner, Reynier Peletier,
Marc Sarzi, and the other \Sauron\ nazguls for their help during the observation campaign, 
as well as for helpful comments. Thanks to Tim de Zeeuw for a detailed reading of the manuscript.
KF acknowledges support provided by NASA through a
grant from the Space Telescope Science Institute, which is operated
by the Association of Universities for Research in Astronomy,
Incorporated, under NASA contract NAS5-26555.
CGM acknowledges financial support from the Royal Society.
This research has made use of the NASA/IPAC Extragalactic Database (NED), 
under contract with the National Aeronautics and Space Administration (NASA). 
This publication makes use of data products of the 2MASS, which is a joint project of 
the Univ. of Massachusetts and the Infrared Processing and Analysis Center, 
funded by the NASA and NSF. We have made use of the LEDA database (http://leda.univ-lyon1.fr).
Based on photographic data obtained using The UK Schmidt Telescope. The UK Schmidt Telescope was operated by the Royal Observatory Edinburgh, with funding from the UK Science and Engineering Research Council, until 1988 June, and thereafter by the Anglo-Australian Observatory. Original plate material is copyright of the Royal Observatory Edinburgh and the Anglo-Australian Observatory. The plates were processed into the present compressed digital form with their permission. The Digitized Sky Survey was produced at the Space Telescope Science Institute under US Government grant NAG W-2166.
Our computations were partly performed on the IBM-SP4
hosted by IDRIS/CNRS and the CRAL 18 nodes cluster funded by the
INSU/CNRS (ATIP \# 2JE014 and Programme National Galaxies).

\label{lastpage}

\end{document}